\documentclass[aps,pre,twocolumn,groupedaddress]{revtex4-2}
\usepackage{amssymb}
\usepackage{amsmath}
\usepackage{color}
\usepackage{sidecap}
\usepackage{graphicx}
\usepackage{bbold}
\usepackage{mathtools}
\usepackage{revsymb}

\usepackage{xcolor,hyperref}
\hypersetup{
   colorlinks,
   linkcolor={blue!50!black},%{red!80!black},
   citecolor={blue!50!black},
   urlcolor={blue!80!black}
}
\DeclareMathAlphabet{\mathitbf}{OML}{cmm}{b}{it}
\newcommand{\dbar}{{\,\mathchar'26\mkern-12mu d}}
\newcommand{\zerovector}{\mathBold 0}

\newcommand{\sFrac}[2]{{\textstyle\frac{#1}{#2}}}

\newcommand{\xv}{\mathitbf x}

\newcommand{\qv}{\mathitbf q}

\newcommand{\uv}{\mathitbf u}
\newcommand{\vv}{\mathitbf v}

\newcommand{\wv}{\mathitbf w}
\newcommand{\zv}{\mathitbf z}

\newcommand{\dv}{\mathitbf d}
\newcommand{\Fv}{\mathitbf F}
\newcommand{\xdot}{\dot{\mathitbf x}}
\newcommand{\xddot}{\ddot{\mathitbf x}}
\newcommand{\xdddot}{\dddot{\mathitbf x}}

\DeclareMathOperator{\Tr}{Tr}
\newcommand{\calBold}[1]{\mbox{\boldmath${\cal #1}$}}
\newcommand{\mathBold}[1]{\mbox{\boldmath$#1$}}
\DeclareMathOperator{\doubleCdot}{\vcentcolon}
\DeclareMathOperator{\quadCdot}{\vcentcolon\hspace*{0em}\vcentcolon}
\DeclareMathOperator{\tripleCdot}{\vcentcolon\hspace*{-0.3em}\cdot}

\begin{document}

\title{Connecting microscopic and mesoscopic mechanics in model structural glasses}
\author{David Richard}
\email{david.richard@ifsttar.fr}
\affiliation{Univ. Grenoble Alpes, CNRS, LIPhy, 38000 Grenoble, France}
\affiliation{Navier, Univ Gustave Eiffel, CNRS, Marne-la-Vallée, France}

\begin{abstract}
We present a novel formalism to characterize elastic heterogeneities in amorphous solids. In particular, we derive high-order strain-energy expansions for pairwise energies under athermal quasistatic dynamics. We then use the presented formalism to study the statistical properties of pairwise expansion coefficients and their link with the statistics of soft, quasilocalized modes, for a wide range of formation histories in both two- and three-dimensional systems. We further exploit the presented framework to access local yield stress maps by performing a non-linear stress-strain expansion within a cavity embedded in a frozen matrix. We show that our "bond micromechanics" compare well with the original "frozen matrix" method, with the caveat of overestimating large stress activations. We additionally show how local yield rules can be used as input for a scalar elasto-plastic model (EPM) to predict the stress response of materials ranging from ductile to brittle. Finally, we highlight some of the limits of simple mesoscale models in capturing the aging dynamics of post-yielding systems. Intriguingly, we observe subdiffusive and diffusive shearband growths for particle-based simulations and EPMs, respectively.
\end{abstract}

\maketitle

\section{Introduction}

Unlike in crystalline materials, where plasticity emanates from dislocations that are well defined topological defects, dissipation in amorphous solids stems from local shear transformation zones (STZs) that lack a clear geometrical signature~\cite{argon1976mechanism,spaepen1977microscopic}. The statistics and collective dissipative dynamics of these microscopic instabilities will control the yielding transition that separates an elastic solid from a flowing state. This transition can exhibit extensive "ductile" flow but also catastrophic "brittle" failure. Here, various emergent phenomena are at play such as scale free avalanches~\cite{antonaglia2014bulk}, necking~\cite{wang2005tensile}, void nucleation and growth~\cite{richard2023bridging}, and shear banding~\cite{vormelker2008effects}. Moreover, the same material can also undergo a so called ductile-to-brittle transition as a function of a control parameter such as the ambient temperature~\cite{lu2003deformation,vormelker2008effects,shi2014intrinsic,richard2023bridging}, strain rate~\cite{lu2003deformation,ketkaew2018mechanical,richard2023bridging}, loading geometry~\cite{richard2021brittle} or even more surprisingly its formation history~\cite{ketkaew2018mechanical}. Modeling the interplay between the density of plastic defects and their interactions is central to understanding the relation between the non-equilibrium history of a material, its structure, and its ability to resist failure for a given loading condition~\cite{shao2020effect,wondraczek2022advancing}. However, mesoscopic (e.g., lattice models~\cite{bulatov1994stochastic,homer2009mesoscale,nicolas2018deformation,van2021implementation}) and macroscopic (e.g., free volume~\cite{spaepen1977microscopic} or shear transformation zone theory~\cite{falk1998dynamics}) models largely rely on untested phenomenological assumptions~\cite{falk1998dynamics,rodney2011modeling,nicolas2018deformation}, such as local flow rules and how the material is renewed after a plastic event.

At the mesoscopic scale, one popular class of models that have been extensively developed are elasto-plastic models (EPMs)~\cite{nicolas2018deformation}. Here, the system is already coarse-grained at the level of the size of an STZ. Each element is loaded elastically until it reaches its own yield criterion (drawn from an initial distribution). After a plastic event, stress is being redistributed across the system in form of an Eshelby-like decay~\cite{eshelby1957determination}. These models have been successful at capturing some of the salient features of amorphous plasticity such as the statistics and dynamics of avalanches post yielding~\cite{liu2016driving,lin2014scaling,ferrero2019criticality,richard2023mechanical}, as well as strain localization across the ductile-to-brittle transition~\cite{talamali2012strain,popovic2018elastoplastic,ozawa2018random,castellanos2022history,xiao2023machine}. Unfortunately, those models rely on many different underlying rules (local dynamics, yield criterion, and renewal distributions) and level of complexity (scalar/tensorial), with a lack of quantitative connection with the actual micromechanics seen in computer glasses or even in real-space laboratory experiments such as in colloidal glasses or foams.

At the microscopic scale, a considerable effort has been made to develop computational methods that are able to measure the degree of ``mechanical disorder" in computer glasses, as well as to estimate local yield stresses that, in turn, feed the formulation of mesoscopic models of plasticity~\cite{patinet2016connecting,richard2020predicting}. Structural indicators encompass purely structural bond order parameters~\cite{malins2013identification,tong2018revealing,paret2020assessing}, machine learning based methods~\cite{ronhovde2011detecting,cubuk2015identifying,boattini2020autonomously,fan2021predicting,xiao2023machine}, the exploration of the potential energy landscape (PEL) of a glassy sample~\cite{rodney2009distribution,tanguy2010vibrational,manning2011vibrational,schwartzman2019anisotropic,kapteijns2020nonlinear,xu2021atomic,richard2021simple,richard2023detecting}, and cavity methods such as the "frozen matrix" that locally shear every part of a configuration~\cite{puosi2015probing,patinet2016connecting,barbot2018local,adhikari2023soft,rottler2023thawed}. Taken together, these indicators decisively showed that disordered solids exhibit a clear signature of micromechanical/elastic heterogeneities with structural motifs that feature a high susceptibility to rearrange under applied stresses. These soft regions have been linked with the presence of soft localized modes associated with the crossing of low energy barriers~\cite{rodney2009distribution,xu2021atomic,kapteijns2020nonlinear}. Mechanically informed metrics have also highlighted that the plastic susceptibility of a given region is a tensorial quantity that depends on the geometry of the remote loading~\cite{barbot2018local,schwartzman2019anisotropic,fan2021predicting}, i.e., one region can soften in pure shear but stiffen under simple shear deformation. Moreover, those methods have been applied in glasses featuring a wide range of mechanical stabilities and could capture a drastic depletion of soft spots for well annealed glasses~\cite{patinet2016connecting,wang2019low,rainone2020pinching,richard2023detecting}.

Cavity methods, such as the original ``frozen matrix" methodology, have also demonstrated that it is possible to extract from a glassy configuration a local yield stress distribution~\cite{puosi2015probing,barbot2018local}. The latter is one of the essential ingredients necessary in order to parametrize EPMs~\cite{nicolas2018deformation}. There have been successful attempts to couple atomistic simulations with lattice models in order to study the yielding transition in the quasistatic limit~\cite{castellanos2021insights,castellanos2022history}, at a constant shear rate~\cite{liu2021elastoplastic}, as well as creep flows at a constant imposed stress~\cite{liu2021elastoplastic}. However, the ``frozen matrix" methodology still requires shearing every part of the sample and thus remains relatively cumbersome. In this work, we propose to test the extent to which a bare as-cast glassy configuration (without any applied deformations) can be informative of the yield stress distribution of a material. Here, we develop a novel framework based on the non-linear strain expansions of the stress at the local level. Borrowing the idea of the "frozen matrix" method, we compute the analytical stress-strain expansion within a cavity embedded within a frozen matrix subjected to an affine deformation. Our framework allows us to approximate local yield stress without having to shear our sample. Further, we show how this method can be used to parametrize a coarse-grained lattice model. We illustrate the mapping between atomistic simulations and a mesoscopic model by predicting the stress-strain response of materials across the ductile-to-brittle transition.

This paper is organized as follows; we first provide in Sec.~\ref{sec:formalism} our new non-linear formalism and derived expressions for local strain derivatives of pairwise energies and stresses. In Sec.~\ref{sec:statistics}, we show the statistics of strain expansion coefficients as a function of material preparation in both 2D and 3D solids. We derive their asymptotic scaling and relation with the underlying population of non-phononic quasilocalized soft modes. We then present in Sec.~\ref{sec:cavity} how this formalism can be used to estimate local yield stresses and compare its limit with the original "frozen matrix" method. Next, a local yield stress distribution is used as an input for a scalar elasto-plastic model to predict stress-strain response for a wide range of glass stability in Sec.~\ref{sec:sim_epm}. We highlight some of the limitations of simple scalar lattice models to capture the correct dynamics of shearband growth during the post-yielding phase. Finally, in Sec.~\ref{sec:discussion} we discuss future work and possible improvements of our framework.

\section{Formalism}
\label{sec:formalism}

Our main goal is to predict how a scalar or tensorial quantities change under an imposed deformation in the athermal, quasistatic limit, where the strain $\gamma$ serves as the key external control parameter~\cite{maloney2004universal,maloney2006amorphous}. In what follows, we will show how to compute high-order expansion coefficients of any observable in terms of the imposed deformation.

In this work, we consider a system of $N$ particles with potential energy $U$, in a box of size $L$, and volume $V\!=\!L^\dbar$, with $\dbar$ standing for the spatial dimension. Tensors are denoted with a boldface font, and contractions over particle indices \emph{and/or} Cartesian components are denoted by dots, e.g.~$\calBold{Q}\!::\!\vv\uv\wv\zv$ denotes a quadruple contraction between the tensor $\calBold{Q}$ and the vectors $\vv,\uv,\wv$ and $\zv$.

\subsection{Strain expansion coefficients}

We consider an observable ${\cal O}(\xv)$ which depends on coordinates $\xv$, and that observable's expansion in terms of a strain parameter, denoted $\gamma$. The latter controls the imposed (affine) transformation $\calBold{D}(\gamma)$, such that coordinates transform as $\xv\!\to\!\calBold{D}(\gamma)\cdot\xv$. As an example, the transformation corresponding to simple shear along the $x$-axis (in 2D) reads
\begin{eqnarray}
\begin{aligned}
\label{eq:simpleshear}
\calBold{D}_{\mbox{\tiny simple shear}}(\gamma)=\left(\begin{array}{cc} 1 & \gamma\\ 0 & 1 \end{array}\right)\,.
\end{aligned}
\end{eqnarray}

In the so-called athermal quasistatic deformation protocol, imposed affine deformations are typically followed by a potential-energy minimization that restores mechanical equilibrium, i.e.~particles undergo additional, \emph{nonaffine} displacements that result in a force-balanced state~\cite{lutsko1989generalized,maloney2006amorphous,lemaitre2006sum}. The total change in coordinates $\xv$ after these aforementioned imposed-deformation and relaxation steps is 
\begin{equation}\label{eq:total_variation}
\frac{d\xv}{d\gamma} = \frac{\partial \xv}{\partial \gamma} + \xdot=\calBold{A}\cdot\xv+\xdot,
\end{equation}
where $\xdot$ denotes the (still unspecified) nonaffine displacements (sometimes referred to as the nonaffine velocities, and see further discussion below), and we define $\calBold{A}\!\equiv\!\partial \calBold{D}/\partial \gamma$ for the sake of brevity. 

With the total variation of coordinates --- as spelled out in Eq.~(\ref{eq:total_variation}) --- at hand, the total derivative of a general observable ${\cal O}$ reads
\begin{equation}\label{eq:derivative_operator}
    \frac{d{\cal O}}{d\gamma} = \frac{\partial{\cal O}}{\partial\xv}\cdot\frac{d\xv}{d\gamma} = \frac{\partial{\cal O}}{\partial\xv}\cdot(\calBold{A}\cdot\xv+\xdot)\,.
\end{equation}

Within the athermal, quasistatic limit, we require that \emph{net} forces on particles $\Fv$ remain unchanged under the imposed deformation, namely $d\Fv/d\gamma\!=\!\zerovector$. Employing Eq.~(\ref{eq:derivative_operator}) above, leads to
\begin{equation}
    \frac{d\Fv}{d\gamma} = \frac{\partial \Fv}{\partial\xv}\cdot(\calBold{A}\cdot\xv+\xdot) = \zerovector\,.
\end{equation}
We identify the Hessian tensor $\partial\Fv/\partial\xv\!\equiv\!-\calBold{H}$ and rearrange the above equation in favor of the nonaffine displacements $\dot{\xv}$, to obtain~\cite{lutsko1989generalized,lemaitre2006sum,hentschel2011athermal}
\begin{equation}
\label{eq:xdot}
\xdot = -\calBold{H}^{-1}\cdot\frac{\partial \Fv}{\partial \gamma},
\end{equation}
where $\partial \Fv/\partial \gamma=\calBold{H}\cdot (\partial \xv/\partial \gamma)$ is the affine shear force. Solving the above equation and plugging $\xdot$ into Eq.~\ref{eq:derivative_operator} allows one to compute the linear strain dynamics of any observable ${\cal O}(\gamma)\simeq {\cal O}_0 + C_\gamma \gamma$, with $C_\gamma=d{\cal O}/ d\gamma|_0$ evaluated at the reference coordinates. We will now generalize this formalism to derive higher-order strain coefficients. We aim to go up to the cubic term $\sim\gamma^3$ in order to probe non-linear asymmetries for between forward and backward deformations. 

The general expansion of ${\cal O}$ with respect to $\gamma$ from a reference coordinate $\xv_0$ takes the form
\begin{eqnarray}
{\cal O}(\gamma)  &\simeq& {\cal O}_0 + \sum_n \frac{1}{n !} C_n \gamma^n 
\end{eqnarray}
with $C_n = d^n {\cal O} /d\gamma^n |_0$, where $|_0$ indicates that derivatives are evaluated on the undeformed coordinates $\xv_0$. In addition we define \emph{total} coordinates derivatives as $\dv_n=d^n\xv/d\gamma^n$, where we have already shown that $\frac{d{\cal O}}{d\gamma}=\partial {\cal O}/\partial\xv \cdot \dv_1$ with $\dv_1 = \calBold{A}\cdot\xv+\xdot$. The second and third coefficients naturally follow as
\begin{eqnarray}\label{eq:second_coef}
\frac{d^2{\cal O}}{d\gamma^2} = \frac{\partial^2 {\cal O}}{\partial\xv\partial\xv}\doubleCdot \dv_1\dv_1 + \frac{\partial {\cal O}}{\partial\xv}\cdot\dv_2,
\end{eqnarray}
and
\begin{eqnarray}\label{eq:third_coef}
\frac{d^3{\cal O}}{d\gamma^3} &=& \frac{\partial^3 {\cal O}}{\partial\xv\partial\xv\partial\xv}\tripleCdot \dv_1\dv_1 \dv_1 + 3\frac{\partial^2 {\cal O}}{\partial\xv\partial\xv}\doubleCdot \dv_1\dv_2 \nonumber \\
&+& \frac{\partial {\cal O}}{\partial\xv}\cdot\dv_3.
\end{eqnarray}

Now, we are missing partial derivatives of a given observable with respect to particle coordinates $\xv$ as well as $\dv_2$ and $\dv_3$. We iteratively take total derivative $d/d\gamma$ on $\dv_1$ and $\dv_2$ to arrive at
\begin{eqnarray}\label{eq:d2}
\dv_2 = \calBold{A}\cdot \frac{d\xv}{d\gamma} + \frac{d\xdot}{d\gamma}= \calBold{A}\cdot\dv_1+\xddot,
\end{eqnarray}
and
\begin{eqnarray}
\dv_3 = \calBold{A}\cdot\dv_2+\xdddot.
\end{eqnarray}
Two unknowns are left in order to evaluate $\dv_2$ and $\dv_3$, namely the non-affine acceleration $\xddot=d\xdot/d\gamma$ and jerk $\xdddot=d\xddot/d\gamma$. The terms \emph{acceleration} and \emph{jerk} are chosen in analogy with time derivatives as $\xdot$ is often called in the community the non-affine \emph{velocity}. Taking total derives $d/d\gamma$ on Eq.~\ref{eq:xdot} one arrives at two new linear equations to solve in favor of $\xddot$ and $\xdddot$ that read
\begin{equation}\label{eq:accel}
\calBold{H}\cdot \xddot = -\Big[\calBold{T}\doubleCdot \dv_1\dv_1 + \calBold{H}\cdot (\calBold{A}\cdot\dv_1)\Big] , 
\end{equation}
and
\begin{equation}\label{eq:jerk}
\calBold{H}\cdot \xdddot = -\Big[\calBold{M}\tripleCdot \dv_1\dv_1\dv_1 + 3\calBold{T}\doubleCdot \dv_1 \dv_2 + \calBold{H}\cdot (\calBold{A}\cdot\dv_2)\Big].
\end{equation}
Here, $\calBold{T}=\partial^3 U/\partial\xv\partial\xv\partial\xv$ and $\calBold{M}=\partial^4 U/\partial\xv\partial\xv\partial\xv\partial\xv$ are the third and fourth derivative of the potential energy with respect to coordinates $\xv$, respectively. Contractions of the different tensors $\calBold{H}$, $\calBold{T}$, and $\calBold{M}$ with vectors are provided in the Appendix for pairwise potentials.

In order to numerically test our expressions and highlight key features of our framework, we prepare glassy samples in both two- and three-dimensions (2D and 3D, respectively) that are composed of a polydisperse mixture of particles interacting via a modified inverse power law repulsive potential~\cite{lerner2019mechanical}. Glass samples are prepared by minimizing the total potential energy of liquid configurations equilibrated at parent temperatures $T_{\rm p}$ using the Swap Monte Carlo (SWAP MC) algorithm~\cite{ninarello2017models}. Details about the model and preparation can be found in the appendix.

\subsection{Potential energy and stress expansion for pairwise interactions}

Having at hand expressions for displacement fields $\dv_n$, we are now left with derivatives of ${\cal O}$ with respect to $\xv$. In the following, we give two examples with both a scalar and a tensorial quantity, namely the total potential energy $U(\xv)$ and the stress tensor $\calBold{\sigma}(\xv)$. We consider particles interacting solely with pairwise interactions as often employed in model glass formers. The potential energy reads $U(\xv)=\sum_\iota \varphi_\iota(\xv_\iota)$, where the subscript $\iota$ labels a pair of particles $i$ and $j$ with energy $\varphi_\iota$ separated by a distance $r_\iota=\sqrt{\xv_\iota\cdot\xv_\iota}$, with $\xv_\iota=\xv_j-\xv_i$. Using Eq.~\ref{eq:second_coef} and Eq.~\ref{eq:third_coef}, the strain expansion for $U$ can be written in a compact form as
\begin{widetext}
\begin{eqnarray}\label{eq:energy_exp}
U(\gamma)\simeq U_0 + \underbrace{\Big[\calBold{J}\cdot\dv_1 \Big]}_{\text{$U_\gamma$}}\gamma + \frac{1}{2}\underbrace{\Big[\calBold{H}\doubleCdot \dv_1\dv_1 +  \calBold{J}\cdot\dv_2\Big]}_{\text{$U_{\gamma\gamma}$}}\gamma^2+ \frac{1}{6}\underbrace{\Big[\calBold{T}\tripleCdot \dv_1\dv_1\dv_1 +  3\calBold{H}\doubleCdot\dv_1\dv_2 + \calBold{J}\cdot\dv_3\Big]}_{\text{$U_{\gamma\gamma\gamma}$}}\gamma^3,
\end{eqnarray}
\end{widetext}
where $\calBold{J}=\partial U /\partial \xv=-\Fv$. As both energies and stresses are additive quantities, expressions derived above and in what follows are valid both at the system level and at the pairwise (bond) level. For example the first coefficient can be computed as
\begin{equation}
U_\gamma = \calBold{J}\cdot\dv_1 = \sum_\iota \underbrace{\frac{\varphi_\iota^{'}}{r_\iota} \xv_\iota \cdot (\calBold{A}\cdot\xv_\iota+\xdot_\iota)}_{\text{$\varphi_\gamma$}},
\end{equation}
where $\varphi_\iota^{'}$ is the first derivative of $\varphi_\iota$ with respect to the pairwise distance $r_\iota$. However, one important difference between coefficients $U_\gamma$ and $\varphi_\gamma$ has to be highlighted. As we consider systems in mechanical equilibrium, terms like $\calBold{J}\cdot\xdot$ present in $\calBold{J}\cdot\dv_n$ will be equal to zero when summed over the whole system. As such, divergence entering in $\xdot$ due to the gapless nature of the non-phononic density of states will be absent, it is the reason why the macroscopic energy and stress are not singular quantities. In contrast, at the bond level singularities of non-affine fields will be present in  $\varphi_\gamma$, $\varphi_{\gamma\gamma}$, ..., which as we will show later in Sec.~\ref{sec:statistics} can be used to measure the density of soft defects and predict regions at the onset of a plastic instability.

We now turn to the stress tensor $\calBold{\sigma}$, which can be written as
\begin{equation}
\calBold{\sigma}(\xv) = \frac{1}{V}\frac{\partial U}{\partial\gamma} = \frac{1}{V}\calBold{J}\cdot (\calBold{B}\cdot \xv),
\end{equation}
where $\calBold{B}\!\equiv\!\hat{\alpha}\hat{\beta}$ is a null tensor where only the $\alpha\beta$ component is equal to unity. Note that for shear deformations, the volume is conserved $V(\gamma)=V_0$ and thus the stress dynamics can be expressed solely through the strain expansion of the observable ${\cal O} = \calBold{J}\cdot (\calBold{B}\cdot \xv)$. Expressions for derivatives and contractions of ${\cal O}$ are provided in the Appendix, plugged into Eq.~\ref{eq:second_coef} and Eq.~\ref{eq:third_coef} allow us to compute the shear dynamics of the shear stress up to $\gamma^3$, as $\calBold{\sigma}(\gamma) \simeq \calBold{\sigma}_0 + \calBold{\sigma}_\gamma \gamma + \frac{1}{2}\calBold{\sigma}_{\gamma\gamma}\gamma^2+\frac{1}{6}\calBold{\sigma}_{\gamma\gamma\gamma}\gamma^3$ with:
\begin{equation}\label{eq:stress_exp}
\calBold{\sigma}_\gamma = \frac{1}{V_0}\Big[\calBold{H}\doubleCdot (\calBold{B}\xv) \dv_1 + \calBold{J}\cdot(\calBold{B}\dv_1) \Big],
\end{equation}
\begin{eqnarray}\label{eq:stress_exp_b}
\calBold{\sigma}_{\gamma\gamma}&=&\frac{1}{V_0}\Big[\calBold{T}\tripleCdot (\calBold{B}\xv)\dv_1\dv_1 +2\calBold{H}\doubleCdot (\calBold{B}\dv_1) \dv_1 \nonumber\\
&+&\calBold{H}\doubleCdot (\calBold{B}\xv) \dv_2+ \calBold{J}\cdot(\calBold{B}\dv_2)\Big],
\end{eqnarray}
and
\begin{eqnarray}\label{eq:stress_exp_c}
\calBold{\sigma}_{\gamma\gamma\gamma} &=& \frac{1}{V_0}\Big[\calBold{M}\quadCdot(\calBold{B}\xv) \dv_1\dv_1\dv_1+3\big[\calBold{T}\tripleCdot (\calBold{B}\dv_1)\dv_1\dv_1 \nonumber\\
&+&\calBold{T}\tripleCdot (\calBold{B}\xv)\dv_1\dv_2 + \calBold{H}\doubleCdot (\calBold{B}\dv_1)\dv_2
+\calBold{H}\doubleCdot (\calBold{B}\dv_2)\dv_1\big] \nonumber\\
&+&\calBold{H}\doubleCdot (\calBold{B}\xv) \dv_3 + \calBold{J}\cdot(\calBold{B}\dv_3) \Big],
\end{eqnarray}
where $\calBold{\sigma}_0$ is the stress component of the undeformed solid. In this expression, the loading geometry is encoded in $\calBold{A}$ (embodied in $\dv_n$, cf.~for example Eq.~\ref{eq:d2}), whereas $\calBold{B}$ controls which of the stress tensor components is being considered. As an example, one can follow the evolution of the pressure for an imposed shear deformation and vice versa. Note however that for deformations where the volume is not conserved $d V/d\gamma \ne 0$, one needs to incorporate extra terms if a solid is prestressed~\cite{barron1965second}. We provide a detailed example in the Appendix that illustrate those subtleties.

\subsection{Connection with macroscopic moduli}

We can connect the first coefficient $\sigma_\gamma$ to the linear elastic properties of the sample, namely the shear modulus $\mu$ and bulk modulus $K$. For a system in mechanical equilibrium, the term $\calBold{J}\cdot(\calBold{B}\dv_1)$ in $\sigma_\gamma$ is zero and one can directly recall the microscopic expression for the shear modulus $\mu$ as 
\begin{equation}
\mu = \sigma_\gamma=\frac{1}{V_0}\Big[\underbrace{\calBold{H}\doubleCdot (\calBold{A}\xv) (\calBold{A}\xv)}_{\text{$\frac{\partial^2 U}{\partial\gamma^2}$}} + \underbrace{\calBold{H}\doubleCdot (\calBold{A}\xv)}_{\text{$-\frac{\partial^2U}{\partial\gamma\partial\xv}$}} \cdot \xdot\Big],
\end{equation}
where $\calBold{B}=\calBold{A}=\hat{x}\hat{y}$ is a null tensor with only one shear component equal to unity, e.g. the $xy$ component for the deformation tensor defined above. From the first and second term, one can easily identify the affine and non-affine part of the modulus, respectively.

For the bulk modulus $K$, this connection is not as straightforward. Let us first define the deformation tensor for a dilation with \emph{true} strain $\eta=\ln(V/V_0)/\dbar$. In 2D, the deformation tensor reads
\begin{eqnarray}
\begin{aligned}
\label{eq:dilation}
\calBold{D}_{\mbox{\tiny dilation}}(\eta)=\left(\begin{array}{cc} e^{\eta} & 0\\ 0 & e^{\eta} \end{array}\right)\,.
\end{aligned}
\end{eqnarray}
Next recall that $K$ can be expressed through the derivatives of the pressure with respect to the volume as $K=-VdP/dV$, where $P=-\Tr{\calBold{\sigma}}/\dbar$. Using chain rules and the definition of the true strain gives us
\begin{equation}
K = -V \frac{dP}{d\eta}\frac{\partial \eta}{\partial V}=\frac{1}{d^2} \frac{d \Tr{\calBold{\sigma}}}{d\eta}.
\end{equation}
When evaluating $d\Tr{\calBold{\sigma}}/d\eta$, one needs to remember that the volume is no longer conserved, and therefore, one also needs to take into account the \emph{affine} dynamics of $V$ with respect to $\eta$. Doing so, we finally arrive at
\begin{equation}
K = \frac{1}{\dbar^2} \Big[ \Tr{\calBold{\sigma}_\gamma} -\dbar\Tr{\calBold{\sigma}_0} \Big].
\end{equation}
Together, our formalism allows us to extract the linear and non-linear elasticity at both the system and pairwise bond level.

\section{Statistical properties of expansion coefficients}
\label{sec:statistics}

%%%%%%%%%%%%%%%%%%%%%%%%%%%%%%%%%%%%%%%%%%%%%%%%%%%%%%%%%%%%%%%%%%%%%%%
\begin{figure}[b!]
  \includegraphics[width = 0.5\textwidth]{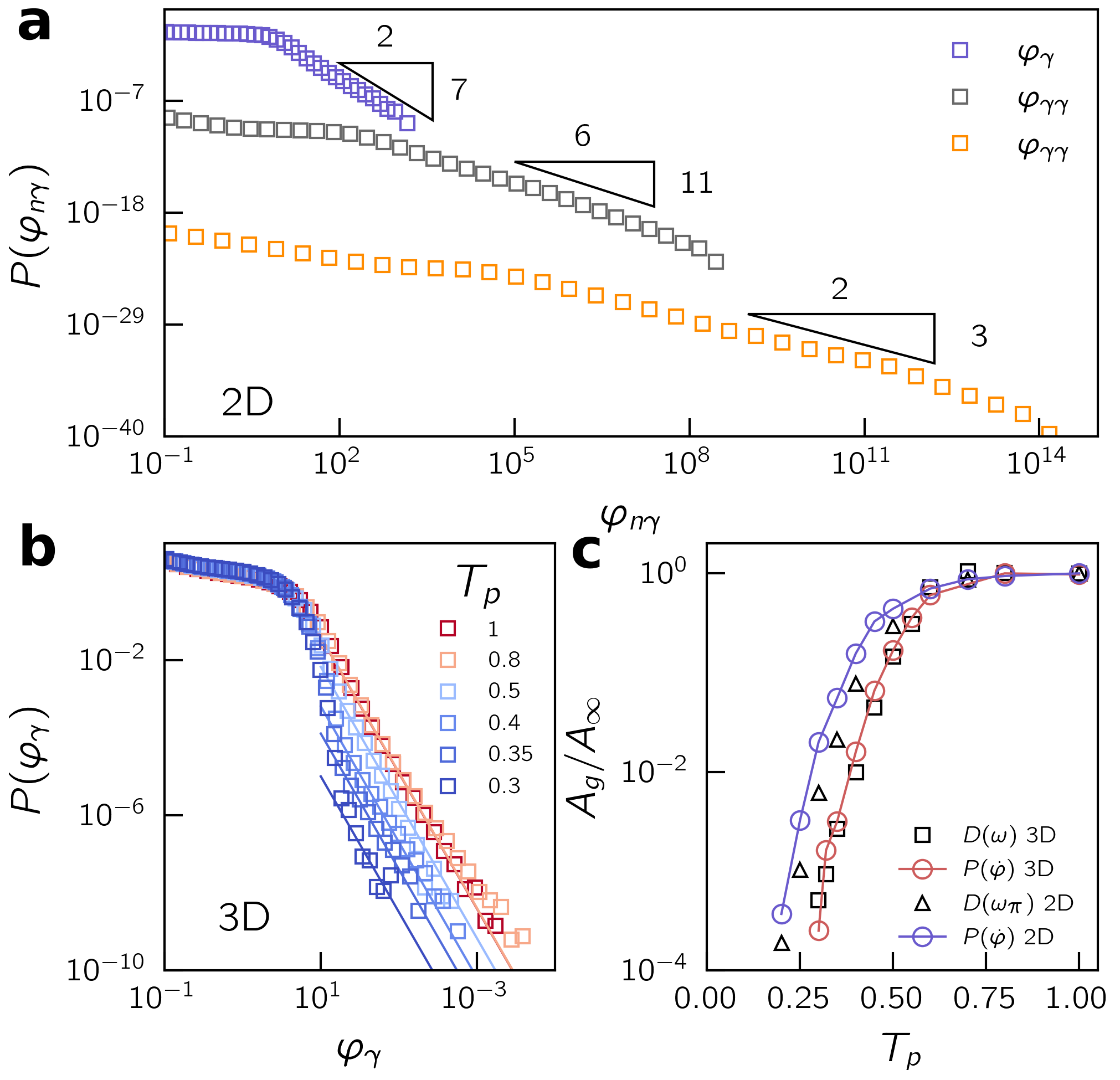}
  \caption{\footnotesize (a) Distributions $P(\varphi_{n\gamma})$ for a 2D poorly annealed samples. (b) $P(\varphi_\gamma)$ in 3D for various parent temperatures $T_p$. Solid lines are fits following $P(\varphi_\gamma)\sim \varphi_\gamma^{-2/7}$. (c) Normalized prefactor of the non phononic density of state $A_{\rm g}/A_\infty$ as a function of $T_p$. Black and triangle squares indicate results from the harmonic and non-harmonic vDOS, respectively. Red and blue circles are results from fitting the asymptotic scaling of $P(\varphi_\gamma)$ for 3D and 2D solids, respectively.}
  \label{fig:stat}
\end{figure}
%%%%%%%%%%%%%%%%%%%%%%%%%%%%%%%%%%%%%%%%%%%%%%%%%%%%%%%%%%%%%%%%%%%%%%%

Here, we provide a discussion and scaling relations that connect properties of strain expansion coefficients at the  bond level with the presence of low-frequency non-phononic excitations coined \emph{quasi-localized modes} (QLMs). The later show a peculiar low-frequency non-phononic density of states $D(\omega)\sim \omega^4$~\cite{lerner2016statistics,mizuno2017continuum}, observed in solid with finite dimensions~\cite{kapteijns2018universal,richard2023detecting}, mean-field models~\cite{bouchbinder2021low}, and for a variety of glass stabilities~\cite{wang2019low,rainone2020pinching,giannini2021bond,richard2023detecting} and models~\cite{richard2020universality}. In both Eq.~\ref{eq:energy_exp} and Eq.~\ref{eq:stress_exp}, singularities in coefficients stem from the presence of the large non-affine components in $\dv_n$, where $n$ indicates the order of the strain derivative. As discussed in Refs.~\cite{maloney2004universal,lerner2016micromechanics}, $\xdot$ in $\dv_1$ is dominated by soft modes present in $\calBold{H}$ that couple well with the shear force $\partial \Fv /\partial \gamma$. A spectral decomposition of Eq.~\ref{eq:xdot} gives the asymptotic relation $\dv_1\sim\xdot\sim \omega^{-2}$ between the non-affine displacement field and the mode frequency $\omega$ of the softest mode. Looking closely at Eq.~\ref{eq:accel} and Eq.~\ref{eq:jerk}, it is easy to generalized such a relation for higher order non-affine terms; one finds $\dv_n\sim\omega^{-\alpha}$, with $\alpha=2+4(n-1)$. Combining this result with the quartic scaling $D(\omega)\sim \omega^4$, one arrives at the following generalized asymptotic scaling
\begin{equation}\label{eq:scaling_qle}
P(C_{n\gamma})\sim C_{n\gamma}^{-\frac{7+4(n-1)}{2+4(n-1)}},
\end{equation}
where the subscript $n\gamma$ stands for the $n$th order coefficient in Eq.~\ref{eq:energy_exp} or Eq.~\ref{eq:stress_exp}.

In Fig.~\ref{fig:stat}(a), we confirm the scaling relation of Eq.~\ref{eq:scaling_qle} for $\varphi_\gamma$, $\varphi_{\gamma\gamma}$, and $\varphi_{\gamma\gamma\gamma}$ in generic bidisperse poorly annealed 2D glasses. Note that $P(\varphi_{n\gamma})$ has to be understood has the distribution of the absolute value of the $n$th order coefficient $\varphi_{n\gamma}=|\varphi_{n\gamma}|$. Next, we investigate the change in the distribution $P(\varphi_\gamma)$ with the preparation protocol. Here, we utilize SWAP MC to equilibrate liquids down to very low parent temperature $T_p$ before a final energy minimization. In Fig.~\ref{fig:stat}(b), we show how $P(\varphi_\gamma)$ varies with $T_p$ for 3D glasses. At low $T_p$, we find a depletion of large $\varphi_\gamma$ values that are associated with regions of large non-affinities. This result is consistent with the aforementioned drop in density of low-frequency quasilocalized modes~\cite{wang2019low,rainone2020pinching}.  We observe the same trend in 2D solids. Note that, the prefactor of the asymptotic tail in $P(\varphi_\gamma)$ has the same unit as the prefactor of $D(\omega)$. In Fig.~\ref{fig:stat}(c), we report the normalized prefactor $A_{\rm g}/A_\infty$ as a function of $T_p$ for both 2D and 3D solids, where $A_\infty$ is the high temperature plateau. We find that $A_{\rm g}$ drops by 3-4 orders of magnitude, consistent with previous results extracted from the harmonic vDOS $D(\omega)$ in 3D~\cite{rainone2020pinching} and anharmonic vDOS $D(\omega_\pi)$ in 2D~\cite{richard2023detecting}, where $\omega$ and $\omega_\pi$ are the frequency of harmonic and (pseudo harmonic) non-linear modes, respectively.

%%%%%%%%%%%%%%%%%%%%%%%%%%%%%%%%%%%%%%%%%%%%%%%%%%%%%%%%%%%%%%%%%%%%%%%
\begin{figure}
  \includegraphics[width = 0.5\textwidth]{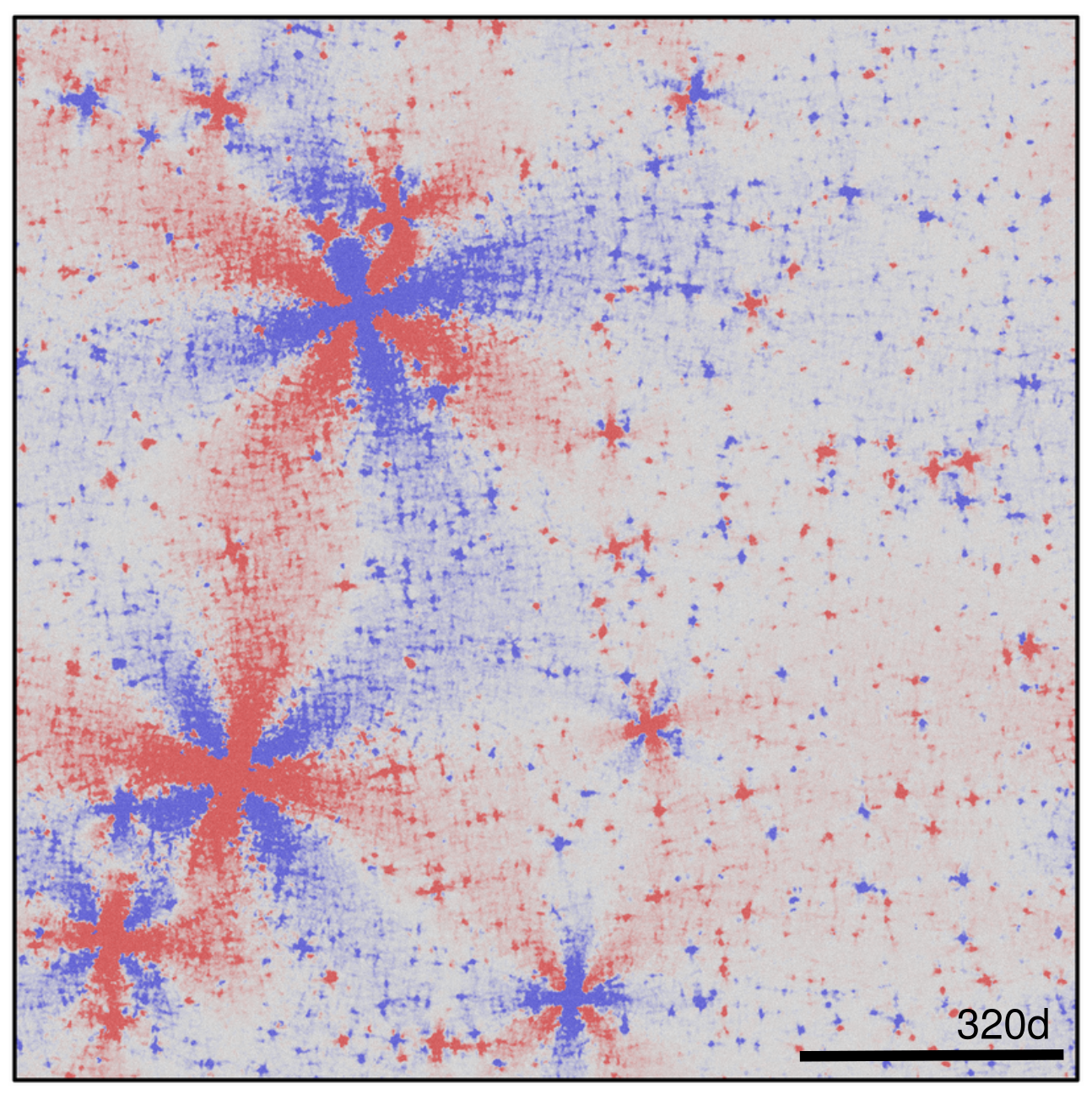}
  \caption{\footnotesize Product field $\varphi_\gamma \times \varphi_{\gamma\gamma}$ for a simple shear deformation along the horizontal direction in a poorly annealed glass with $N=1638400$. Positive (blue) and negative (red) products correspond to regions that will destabilize and stabilize along the forward direction, respectively.}
  \label{fig:product}
\end{figure}
%%%%%%%%%%%%%%%%%%%%%%%%%%%%%%%%%%%%%%%%%%%%%%%%%%%%%%%%%%%%%%%%%%%%%%%

One advantage of our formalism is that measuring all pairwise strain expansion coefficients only require iteratively solving linear equations. Thus, once can quickly visualize pairwise bond field in systems composed of a few million of particles. Moreover, we now have access to both the first and second derivatives of pairwise energy $\varphi_\gamma$ and $\varphi_{\gamma\gamma}$, respectively. In Ref.~\cite{schwartzman2019anisotropic}, authors derived the strain derivatives of the local heat capacity $c_\alpha$ of a pair $\alpha$. They demonstrated that one can construct an anisotropic indicator with the product $c_\alpha \times d c_\alpha/d\gamma$, which allows to predict for which loading geometry and direction a soft spot is likely to be activated. Regions that are likely to be destabilized for small strain increments will show both large compression and decompression of pairwise distances, such as a T1 transition seen in foams~\cite{dennin2004statistics}. Hence, we expect the strain dynamics of pairwise energy $\varphi(\gamma)$ to either decrease or increase a lot with $\gamma$ and thus the product $\varphi_\gamma \times \varphi_{\gamma\gamma}$ to be positive. In contrast, negative product will correspond to a stabilization of the region at order $\mathcal{O}(\gamma^2)$ in the forward loading direction or destabilization in the backward direction. As an illustration we coarse grain the product field $\varphi_\gamma \times \varphi_{\gamma\gamma}$ for each particle by summing all product values within interacting neighbors. In Fig.~\ref{fig:product}, we display such a structural field for a poorly annealed glass with $N=1638400$, where positive and negative product are rendered in blue and red, respectively. This method allows to quickly detect soft spots that couple well with the imposed loading direction and geometry. One clear drawback comes from the singularity of elastic properties where $\varphi_\gamma \times \varphi_{\gamma\gamma}$ diverges in regions where mode's frequency $\omega\to0$, as realized by very large Eshelby-like fields in Fig.~\ref{fig:product}. In practice, such a spatial indicator is a good predictor for the first plastic events occurring within small deformations ($\gamma<1\%$). However, stiffer regions that will soften and activate further in strain are being buried under the halo created by rare low-frequency QLMs. This is the reason why ranked metrics constructed from solely the non-affine velocity show a good correlation with plastic events only within $1-2\%$ of strain~\cite{xu2021atomic}.

\section{Non-linear micromechanics in a "frozen matrix"}
\label{sec:cavity}

%%%%%%%%%%%%%%%%%%%%%%%%%%%%%%%%%%%%%%%%%%%%%%%%%%%%%%%%%%%%%%%%%%%%%%%
\begin{figure}[t!]
  \includegraphics[width = 0.5\textwidth]{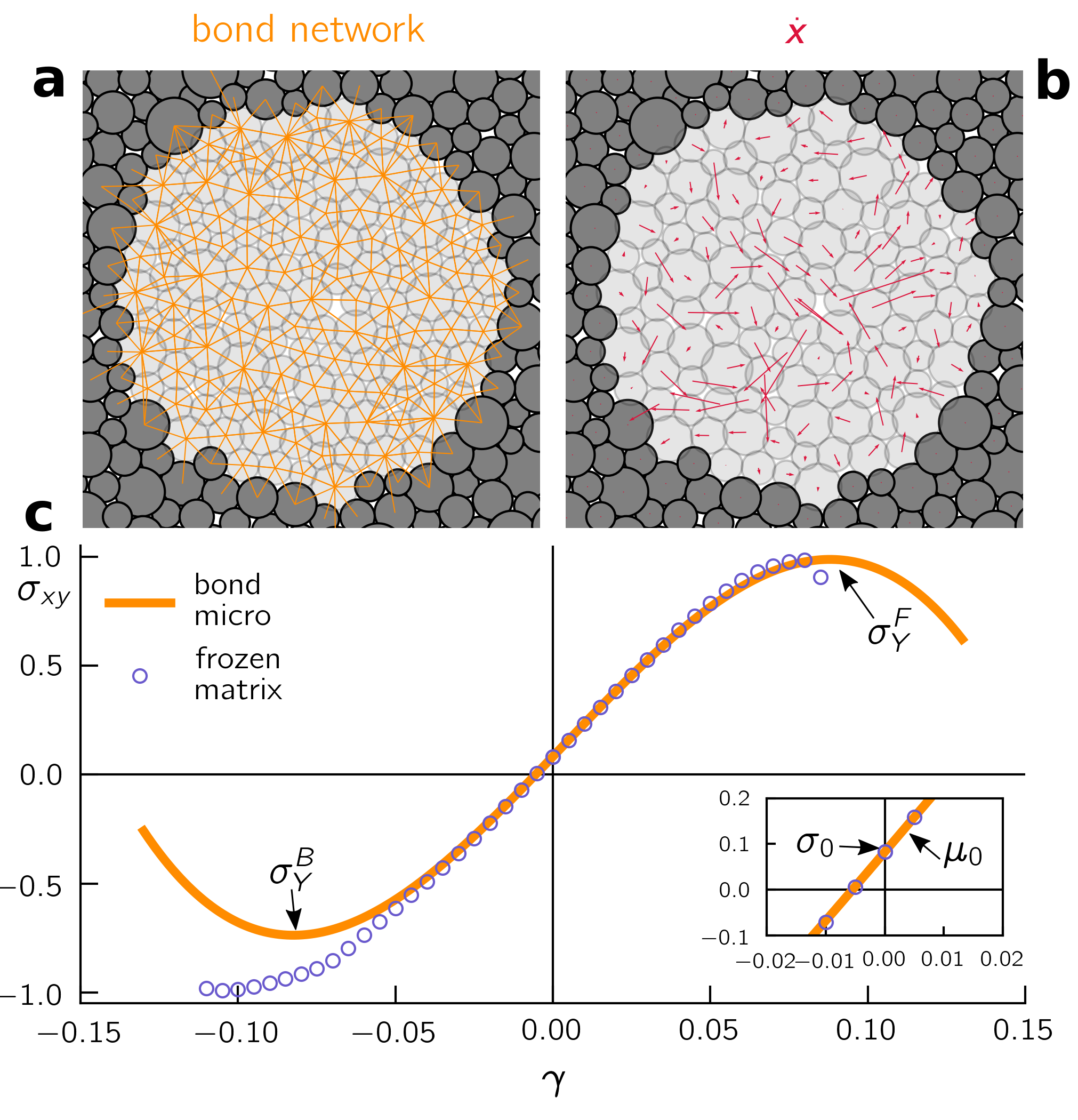}
  \caption{\footnotesize (a) Bond network of pairwise interactions inside a cavity embedded in a frozen matrix (dark particles). (b) Analytical non-affine response inside the cavity for a simple shear deformation. (c) Non-linear stress expansion up to $\gamma^3$ compared with the frozen matrix method. The inset shows a zoom of the linear response with prestress $\sigma_0$ and shear modulus $\mu_0$. $\sigma_Y^{F}$ and $\sigma_Y^B$ indicate the forward and backward yield stress for a simple shear deformation along the x-axis.}
  \label{fig:frozenmat}
\end{figure}
%%%%%%%%%%%%%%%%%%%%%%%%%%%%%%%%%%%%%%%%%%%%%%%%%%%%%%%%%%%%%%%%%%%%%%%

In this section, we propose to use our non-linear formalism to extract estimates for local stress activation thresholds. Here, we will borrow the same idea as in the ``frozen matrix" method~\cite{puosi2015probing,patinet2016connecting}, where a local cavity embedded into a rigid matrix is sheared until a plastic event occurs. First, we select pairs of interacting particles (referred to as a ``bond") that are located within a radius $R$ of one particle embedded into a rigid matrix, see snapshot in Fig.~\ref{fig:frozenmat}(a). Using the reduced Hessian associated with those free particles, we can compute the analytical non-affine response inside the cavity subject to an affine deformation of the frozen part. In Fig.~\ref{fig:frozenmat}(b), we show such a field $\xdot$ for a simple shear deformation. Subsequently, we iteratively solve high order non-affine responses, which can then be plugged into Eq.~\ref{eq:stress_exp}, Eq.~\ref{eq:stress_exp_b}, and Eq.~\ref{eq:stress_exp_c} to compute the non-linear strain expansion of the full stress tensor inside the cavity. In Fig.~\ref{fig:frozenmat}(c), we show the $xy$ component of $\calBold{\sigma}$ plotted as a function of the strain $\gamma$ and compare it with the result obtained by the explicit ``Frozen matrix" method, including the deformation in both backward and forward shear directions. In the original ``frozen matrix" method, we have performed a local athermal quasistatic shear (AQS) deformation with a step $\Delta\gamma=10^{-3}$ until a stress drop is detected. Overall, we find a good semi-quantitative agreement with some deviations at large strains ($\gamma>0.05$) and we recover the same prestress and local shear modulus, as shown in the inset. Deviations between the stress-strain expansion and the true strain dynamics are to be expected due to 3 reasons, namely: (i) the limited order of the expansion, (ii) the convergence radius of the expansion due to singular high order terms, and (iii) possible non-continuous contact change during large deformation prior to yielding~\cite{morse2020differences}. The main advantage of our method is the computational cost as we only require to solve iteratively 3 linear equations without any deformation of the as-cast structure. For example, we can compute an estimate for the complete map of the local yield stress in systems composed of few tens of thousands of particles in a few seconds on a single CPU-core. As a result, one can efficiently gather average statistics over sample fluctuations as well as to monitor the dynamics of the yield stress during deformation.

%%%%%%%%%%%%%%%%%%%%%%%%%%%%%%%%%%%%%%%%%%%%%%%%%%%%%%%%%%%%%%%%%%%%%%%
\begin{figure*}[t!]
\center
  \includegraphics[width = 1\textwidth]{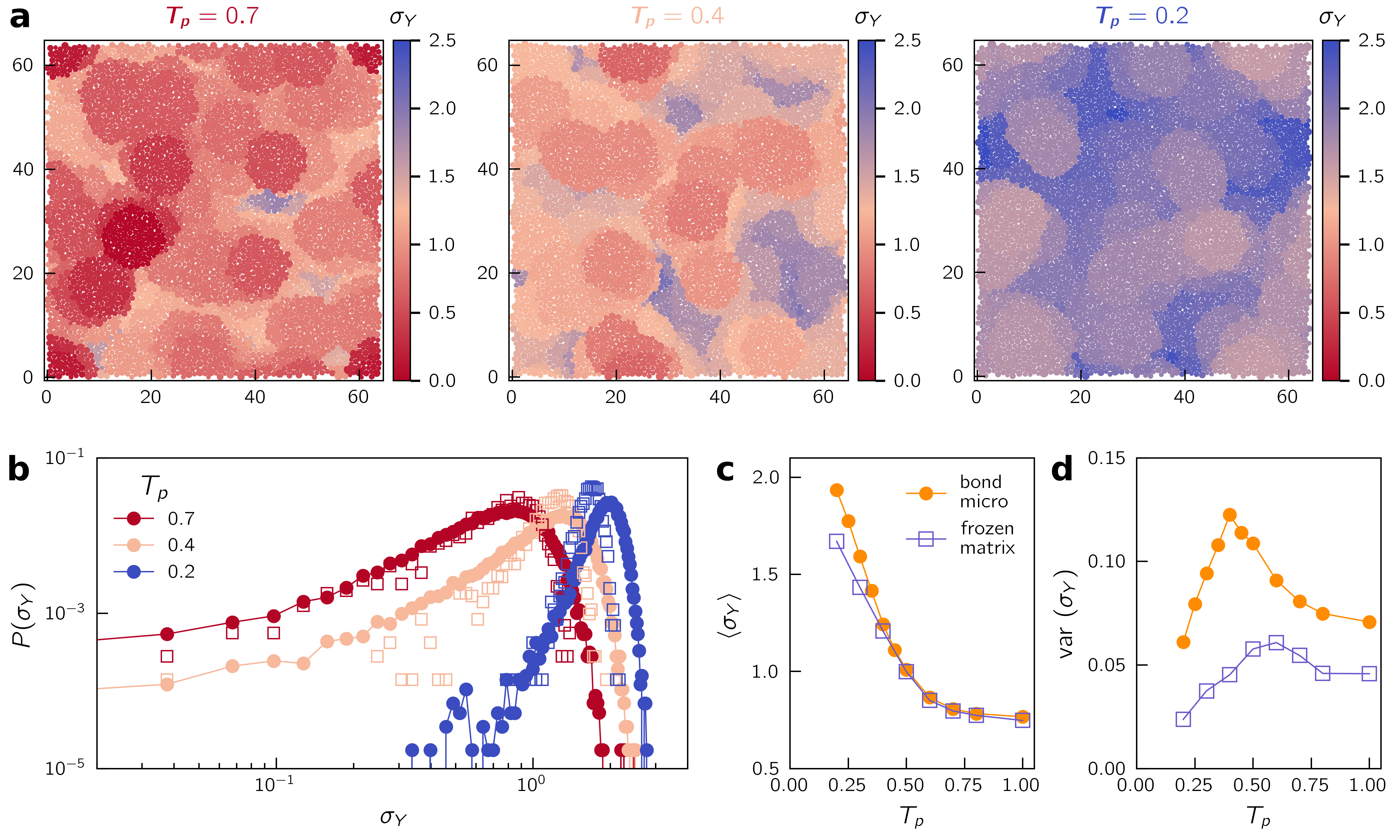}
  \caption{\footnotesize (a) Colormap of the local yield stress extracted in glasses quenched from liquid equilibrated at different parent temperatures: $T_p=0.7$ (left, poorly annealed), $T_p=0.4$ (middle, mildly annealed), and $T_p=0.2$ (right, very stable glass). (b) Distribution of the local yield stress extracted from the bond micromechanics (filled symbols) and frozen matrix (empty symbols) for different degrees of annealing. (c) Average local yield stress versus the parent temperature $T_p$. (d) Variance of the local yield stress against $T_p$.}
  \label{fig:lys_map}
\end{figure*}
%%%%%%%%%%%%%%%%%%%%%%%%%%%%%%%%%%%%%%%%%%%%%%%%%%%%%%%%%%%%%%%%%%%%%%%

We can now apply this method to compute the local yield stress $\sigma_Y$ for any given deformation. In what follows, we select equally spaced particles by $3$ diameters $d=\rho^{-1/\dbar}$, with the number density $\rho=N/V$. For each particle, we perform our bond micromechanics within a radius $R=5d$ and extract the forward yield stress, denoted in what follows by $\sigma_Y$. In the "frozen matrix" method, the system will always eventually yield for sufficiently large strains, meaning that we can always define $\sigma_Y$ for any trial. In contrast, in the "bond micromechanics" framework, there are rare events where the imposed rigid matrix stabilizes the region. For example, this can happen when the rigid interface is placed directly at the core of a plastic instability. As a result, non-linear stress terms can be all positive, translating into a diverging stress function. Those trials are being discarded and particles always receive the lowest $\sigma_Y$ value of all attempts that have been performed in their vicinity. We have checked that with our particle spacing and cavity radius, one always recovers a yield stress value for every particle. 

As an illustration, we propose to apply our methods to glasses exhibiting a wide range of mechanical stability utilizing the same SWAP MC as mentioned and used in Sec.~\ref{sec:statistics}. In Fig.~\ref{fig:lys_map}(a), we show the $\sigma_Y$ map for a forward simple shear deformation in glasses equilibrated at different parent temperatures $T_p$. We observe a strong spatial heterogeneity of $\sigma_Y$ that strongly depend on $T_p$. As the sample is more and more stable, lowering $T_p$ (from left to right), we observe a depletion of regions with a low local yield stress. This observation is fully consistent with previous numerical studies using the "frozen matrix" method~\cite{patinet2016connecting,barbot2018local,richard2020predicting} and in agreement with the depletion of soft localized modes seen in Fig.~\ref{fig:stat} and Refs.~\cite{wang2019low,rainone2020pinching,richard2023detecting}. The overall shift of $\sigma_Y$ to larger values as $T_p$ decreases is already a microscopic indication of an increase of the macroscopic yield stress of the material. This result is remarkable as we only harvest the as-cast configuration.

Next, we quantitatively compare yield stresses obtained with both our "bond micromechanics" and the "frozen matrix". For both methods, we coarse grain our map on a length scale $\xi=5d$, and compute the probability distribution $P(\sigma_Y)$. $\xi$ is set according to the typical core size of a shear deformation as demonstrated in Refs.~\cite{rainone2020pinching,giannini2021bond,richard2023detecting}. This choice is motivated with the aim of using $P(\sigma_Y)$ to calibrate a mesoscale lattice model in the next section. In Fig.~\ref{fig:lys_map}(b), we show the yield stress distribution for various parent temperatures that span the range from poorly annealed to very stable glasses, the same as in Fig.~\ref{fig:lys_map}(a). For both methods, we find a shift of $P(\sigma_Y)$ to larger values as $T_p$ decreases. Furthermore, we observe a sharp depletion of the asymptotic tail $\sigma_Y\to 0$ as the system is more and more annealed. This results is consistent with previous observations~\cite{patinet2016connecting,richard2020predicting}. We then compare the mean and variance of $P(\sigma_Y)$ as a function of $T_p$, see Fig.~\ref{fig:lys_map}(c) and (d), respectively. Our bond micromechanics method is able to recover the same average yield stress of the material as with the "frozen matrix" method, with some small deviations for the lowest $T_p$ values. However, we find that our "bond micromechanics" approach overestimate the variance of the distribution by at least 50\% for the whole range of glass stability. Deviations between the true strain dynamics and non-linear expansion from the as-cast configurations are linked with regions having large barriers. Indeed, looking at the distributions in Fig.~\ref{fig:lys_map}(b), we clearly observe that the strain expansion perform much better in regions close to an instability, but overestimate large yield stresses. Regardless, this result demonstrates that one can extract a meaningful scale for the yield stress in an as-cast configuration without deforming the solid.

\section{Mapping between micromechanics and mesoscopic elasto-plastic models}

\subsection{Stress-strain response}
\label{sec:sim_epm}

One of the main advantages of our approach and the ``frozen matrix" method compared with other indicators is to provide stress activation thresholds as input for mesoscopic models of plasticity. Here, we propose to compare how the different yield stress distributions extracted in Sec.~\ref{sec:cavity} perform as inputs for EPMs in order to predict the stress-strain response of a material. To compare the sole effect of $P(\sigma_Y)$ on results, we keep our EPM as simple as possible. We adopt a scalar quasistatic EPM with synchronous dynamics that have been previously employed in Refs.~\cite{Khirallah2021yielding,richard2023mechanical}. Our model consists of blocks of size $\xi=5d$ (typical core size of an STZ). The system has an initial shear modulus $\mu(T_p)$ taken from the bulk sample equilibrated at $T_p$. The system is elastically loaded by $\mu\gamma$ until the stress $\sigma_i$ of a block $i$ reaches its yield stress $\sigma_Y^i$. After a plastic event, stress is being instantaneously redistributed across the system. Here, we employ Fast Fourier Transforms to speed up the stress update on a square grid. In Fourier space the stress increment reads $\Delta\sigma(\qv)=2\mu\epsilon_{\rm pl}G(\qv)$~\cite{picard2004elastic}, with $G(\qv=\zerovector)=-1$ and
\begin{equation}
G(q_x,q_y)=-4 \frac{q_x^2q_y^2}{q^4},
\end{equation}
for $\qv\ne\zerovector$, with 2D wave vector $\qv=\{q_x,q_y\}$ and square norm $q^2=q_x^2+q_y^2$. Our only adjustable parameter will be the plastic eigenstrain $\epsilon_{\rm pl}$. We consider two distributions $P_{\rm init}$ and $P_{\rm renewal}$ that describe the initial and renewal yield stress distribution, respectively. $P_{\rm init}$ corresponds $P(\sigma_Y,T_p)$ at the parent temperature $T_p$ where the system has been prepared, whereas $P_{\rm renewal}$ is set to the yield stress distribution of a sample quenched from a high temperature (here $P(\sigma_Y,T_p=0.7)$). In other words, our assumption is that once a block fails it directly rejuvenates in the statistics of a poorly annealed glass with a lower average yield stress~\cite{barbot2020rejuvenation}. This approach is different from the one from Ref.~\cite{castellanos2022history} where a mechanical aging time scale that depends on the state of the system was introduced. In this study, authors employed a binary mixture of small and larger particles that can demix at a high temperature. Hence, the aging time scale is likely to reflect a structural remixing of the mixture under strain~\footnote{Sylvain Patinet private communications}. This structural mechanism is completely absent in our highly polydisperse glass former. Additionally, we update the local shear modulus of the block to $\mu(T_p=0.7)$ after a plastic event. The global shear modulus is simply $\mu=\sum_i\mu_i$, where $\mu_i$ is the shear modulus of the block $i$.

%%%%%%%%%%%%%%%%%%%%%%%%%%%%%%%%%%%%%%%%%%%%%%%%%%%%%%%%%%%%%%%%%%%%%%%
\begin{figure}[t!]
  \includegraphics[width = 0.5\textwidth]{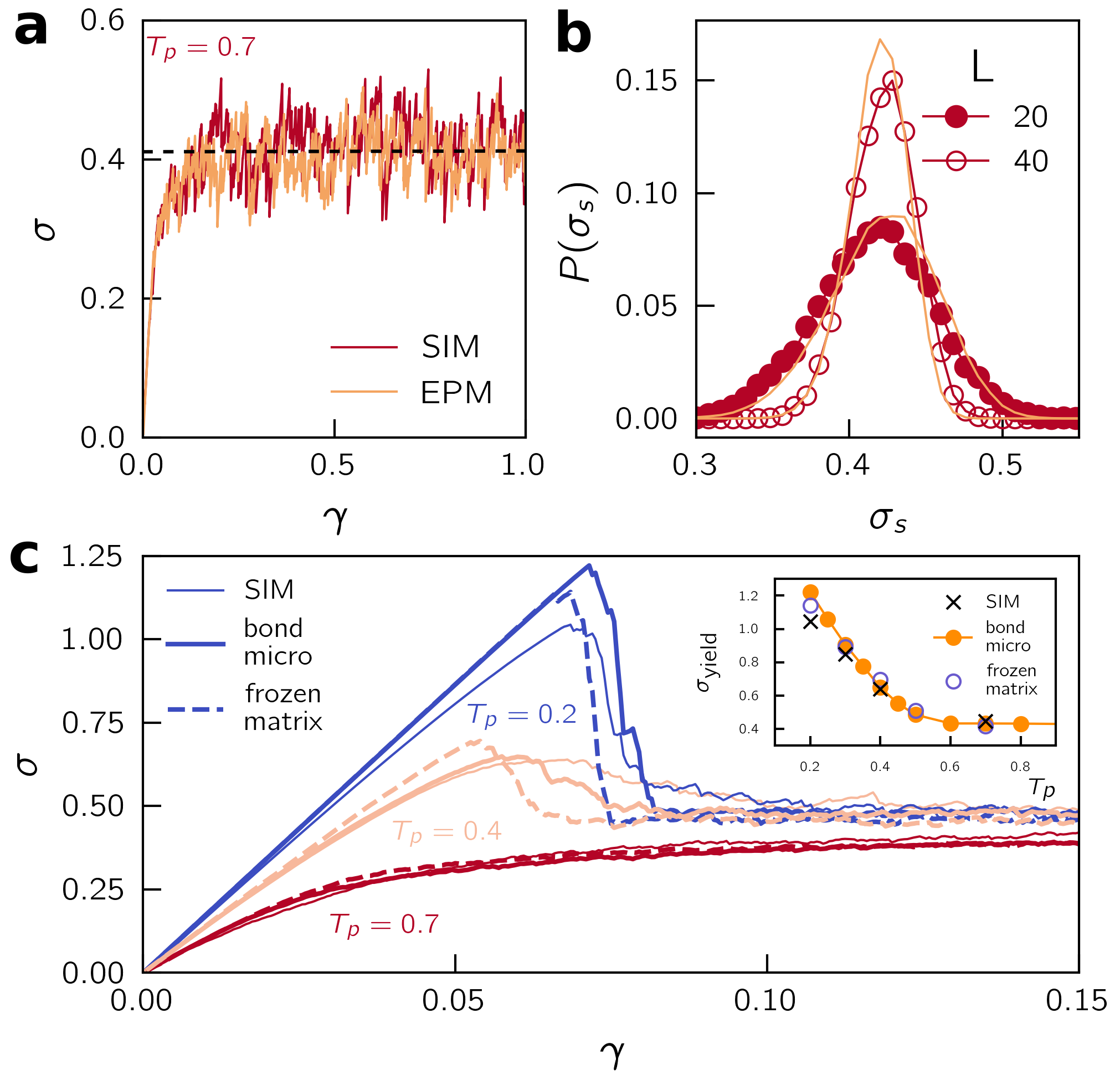}
  \caption{\footnotesize (a) Stress versus strain in particle simulation (red) and elasto-plastic model (orange) for a poorly annealed glass ($T_p=0.7$). The horizontal dash line indicates the average steady state stress $\sigma_s$. (b) Probability distribution of the steady state stress $\sigma_s$ for two different system sizes $L=20$ ($L=100d$) and $L=40$ ($L=200d$), respectively. (c) Stress versus strain for different parent temperatures. Thin solid lines, thick solid lines, and dashed lines correspond to particle simulations, EPM parameterized from the bond micromechanics, and EPM parameterized from the frozen matrix method, respectively. The inset shows the (peak) yield stress as a function of $T_p$.}
  \label{fig:epm}
\end{figure}
%%%%%%%%%%%%%%%%%%%%%%%%%%%%%%%%%%%%%%%%%%%%%%%%%%%%%%%%%%%%%%%%%%%%%%%

In this model, $\epsilon_{\rm pl}$ is the only adjustable parameter and is tuned in order to recover the steady state stress $\sigma_s$ of a poorly annealed sample. In what follows, we have set $\epsilon_{\rm pl}\simeq0.065$ and have have kept it fixed for all sample preparations. Note that in realistic systems, $\epsilon_{\rm pl}$ fluctuates from one STZ to another, see detailed discussions in Refs.~\cite{albaret2016mapping,nicolas2018orientation,castellanos2021insights}. In Fig.~\ref{fig:epm}(a), we compare the stress versus strain curve of one glass realization between atomistic simulations and our EPM for a poorly annealed glass ($T_p=0.7$). Our microscopic system is composed of $N=10000$ particles equivalently to a $20\times 20$ EPM grid. We have also check that a fixed value for $\epsilon_{\rm pl}$ allows not only to recover the correct $\sigma_s$ but also its fluctuations and system size dependence. In Fig.~\ref{fig:epm}(b), we show the stress distribution $P(\sigma_s)$ within the steady state obtained for $2$ different system sizes $L=20$ ($N=10000$) and $L=40$ ($N=40000$). We find a very good agreement between atomistic simulations and our mesoscale model. This result confirms that one can keep an EPM as minimal as possible and still capture relatively well system size dependence stress fluctuations.

We are now in position to see the influence of $P_{\rm init}$ on the material response and compare our "bond micromechanics" and the "frozen matrix" method on the same footprint. In Fig.~\ref{fig:epm}(c), we show $\sigma$ against $\gamma$ for $L=40$ and $3$ different parent temperatures: $T_p=0.7$ (poorly annealed), $0.4$ (mildly annealed), and $0.2$ (very stable). Stress signals are average over different realizations of the initial disorder. Overall, we find a good agreement between atomistic calculations and EPM results. Our mesoscale model is able to capture the increase of both the yield stress $\sigma_{\rm yield}$ and stress drop when $T_p$ is lowered. As shown in the inset, the agreement is very good, only minor deviations are observed for the lowest temperature. This small discrepancy can be linked to the absence of non-linearity in the EPM, i.e., a \emph{reversible} softening that is not due to plastic dissipation but due to non-affinity that enters in the non-linear elastic coefficients. The latter is not negligible for large deformations $\gamma>5\%$. We also find that $\sigma_{\rm yield}$ is better captured by the "frozen matrix" for $T_p=0.2$, which is consistent with the overestimation of large barriers by the bond micromechanics, as seen before in Fig.~\ref{fig:lys_map}(c). Yet, it is fair to conclude that it is possible to semi-quantitatively parameterize an EPM solely from the as-cast solid. In addition, we demonstrate that the instantaneous rejuvenation after one plastic event is good assumption if no other structural rearrangement are at play.

\subsection{Shearband nucleation and growth}
\label{sec:shearband}

%%%%%%%%%%%%%%%%%%%%%%%%%%%%%%%%%%%%%%%%%%%%%%%%%%%%%%%%%%%%%%%%%%%%%%%
\begin{figure*}[t!]
  \includegraphics[width = \textwidth]{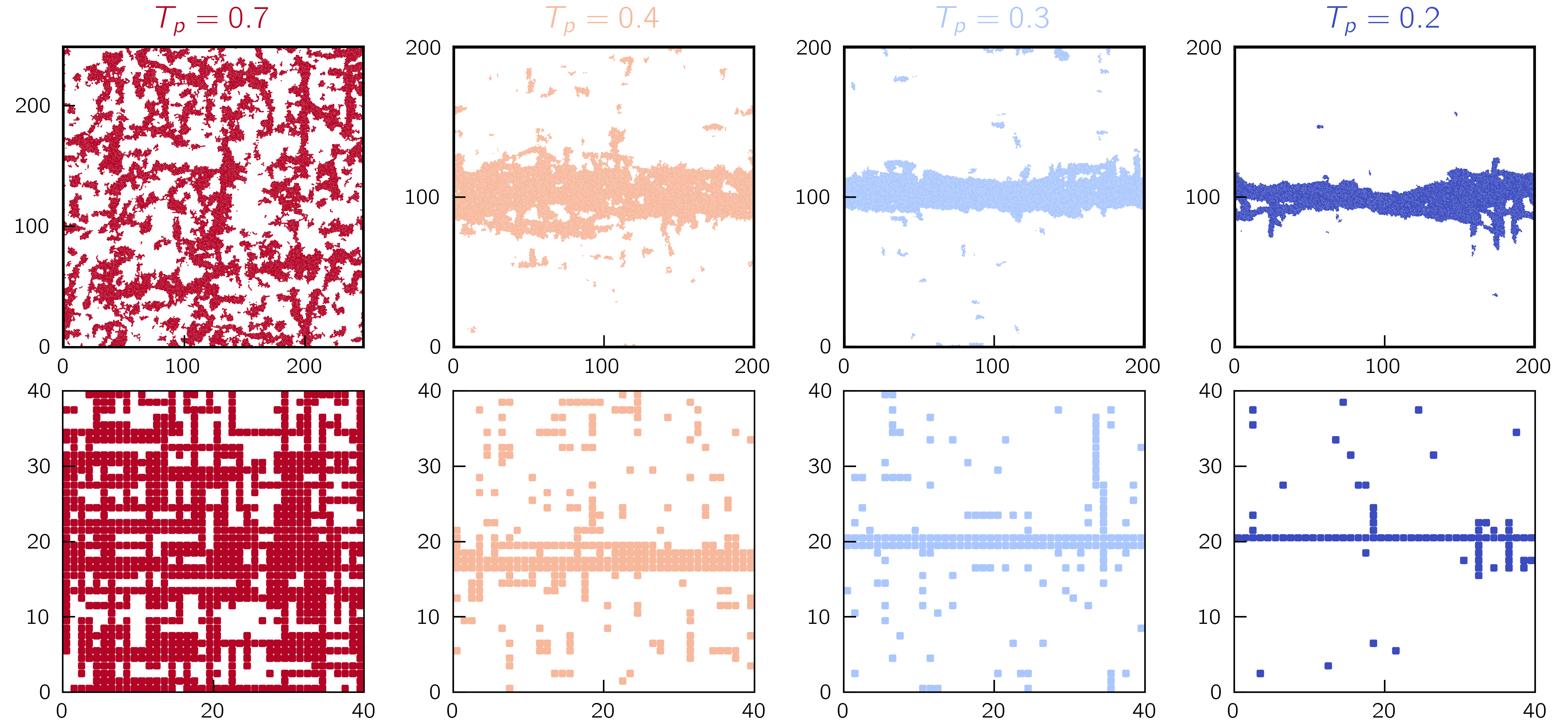}
  \caption{\footnotesize Accumulated plasticity map across the yielding transition from high (left) to low (right) parent temperatures. Simulations and elasto-plastic models are displayed on top and bottom, respectively. Colored particles have yielded at least once. The lateral size of a block in the EPM corresponds to 5 particle diameters.}
  \label{fig:plasticmap}
\end{figure*}
%%%%%%%%%%%%%%%%%%%%%%%%%%%%%%%%%%%%%%%%%%%%%%%%%%%%%%%%%%%%%%%%%%%%%%%

As the size of our EPM is calibrated to match the size of our atomistic box, we can fairly compare the cumulative plastic activity across the yield transition as a function of $T_p$, see Fig.~\ref{fig:plasticmap}. In the EPM, the cumulative plasticity of a given site is accessible directly as we can keep track of the activation within each block. In particle based simulations it is more ambiguous, here we monitor the plastic activity from the $D^2_{\rm min}$ non-affine field~\cite{falk1998dynamics} every $1\%$ of incremental strain. Each particle is considered to have undergone a plastic event for $D^2_{\rm min}>0.01$. Particles or block that have been active within $\gamma<12\%$ are rendered in color in Fig.~\ref{fig:plasticmap}. At a high temperature $T_p=0.7$, we find a rather homogeneous plastic deformation, which is typical for ductile material without a stress overshoot~\cite{ozawa2018random,rossi2023far}. Lowering $T_p$, we observe the appearance of strain localization with the formation of a shearband (SB). The width $w$ of the nucleating SB decreases with deeper annealing. Across the temperature range $T_p=0.4$ to $0.2$, we find $w$ moving from typically $3-4$ to $1$ block in good agreement with atomistic simulations with widths ranging from $30$ to $10$ particle diameters. However, it is important to point out that as the macroscopic stress drop is independent of the system size (in the limit $L\to\infty$), we expect $w$ to grow with the box length $L$ with some power that reflects the spatial spreading of the extensive mechanical unloading during the nucleation process. We show in the Appendix the sub-extensive scaling $w\sim L^{\alpha}$, with $\alpha$ within $0.55-0.75$. The range of $L$ is of course rather limited in our particle based simulations, nonetheless our scaling is surprisingly close the exponent $\alpha\simeq 0.68$ extracted very accurately in similar scalar EPM~\cite{rossi2023far}.

%%%%%%%%%%%%%%%%%%%%%%%%%%%%%%%%%%%%%%%%%%%%%%%%%%%%%%%%%%%%%%%%%%%%%%%
\begin{figure}[b!]
  \includegraphics[width = 0.5\textwidth]{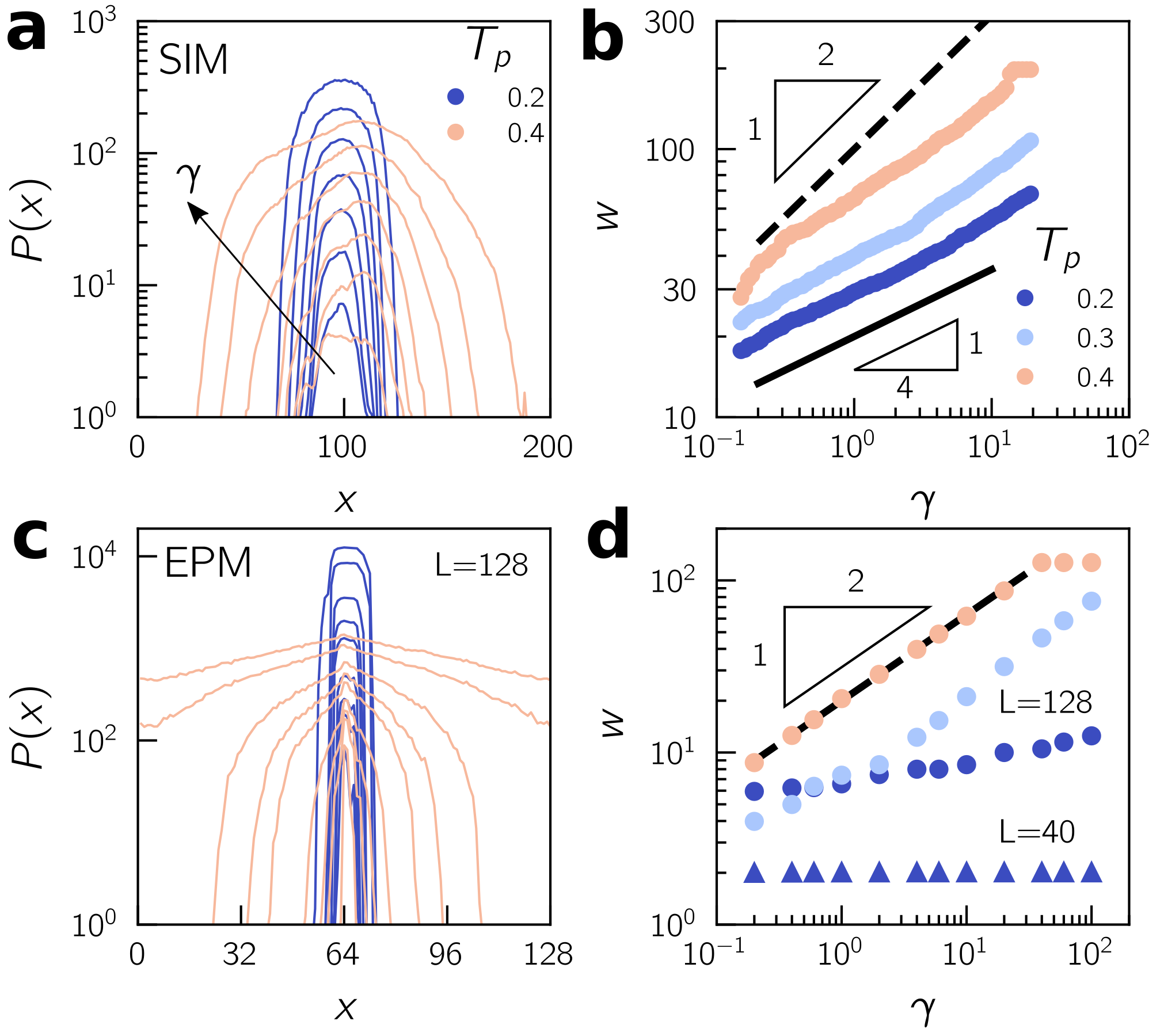}
  \caption{\footnotesize (a) Particle based simulations plasticity profile centered at the position of the shearband for different increments of strain $\gamma=0.18$, $0.36$, $0.72$, $1.43$, $2.83$, $5.61$, and $11.12$. Light pink and blue correspond to a mildly annealed and very stable glass, respectively. (b) Width of the profile as a function of the strain for different parent temperatures. (c) and (c) are the same as in (a) and (b) for the elasto-plastic model parametrized with our bond micromechanics. Increments of strain in (c) are $\gamma=0.2$, $0.6$, $1$, $2$, $6$, $10$, $20$, $60$, and $100$. Circles and triangles in (d) are for $L=128$ and $L=40$, respectively.}
  \label{fig:sbgrowth}
\end{figure}
%%%%%%%%%%%%%%%%%%%%%%%%%%%%%%%%%%%%%%%%%%%%%%%%%%%%%%%%%%%%%%%%%%%%%%%

As a last comparison, we propose to follow the post-yielding dynamics of shearband for different material preparation. We define the plastic activity $P$ as the number of shear event that a particle or a EPM block have experienced. Having located the position of the SB, we can then track the plastic profile $P(x)$, with $x$ being the position along the transverse direction of the SB. Here, we consider only the case of a single shearband that nucleates along the horizontal direction, avoiding the turnover of a vertical SB during large deformations, see detailed discussion in Ref.~\cite{alix2018shear}. In Fig.~\ref{fig:sbgrowth}(a), we show $P(x)$ for atomistic simulations with $N=40000$ at different strain amplitude $\gamma$ and $2$ different parent temperatures, corresponding to a mildly ($T_p=0.4$) and extremely well ($T_p=0.2$) annealed glasses. For both material preparation, we observe a progressive invasion of the shearband in whole system. For the same applied strain, we find that the width $w$ to be significantly smaller for the most aged sample ($T_p=0.2$). Note that we are not able to observe the complete shear melting of the system even for strains up to $2000\%$. In Fig.~\ref{fig:sbgrowth}(b), we plot $w$ as a function of $\gamma$. Interestingly, we observe a subdiffusive growth for all $T_p$ with $w\sim \gamma^{0.25}$. This power law breaks down when $w$ starts to approach the size of the system, where the SB grows faster. This result is consistent with 3D subdiffusive shearband growth observed in Ref.~\cite{golkia2020flow}, where authors found $w\sim\gamma^{0.32}$. Note that we do observe a diffusive growth with $w\sim\sqrt{\gamma}$ in smaller systems $N=10000$ and $N=2500$ that are mildly annealed ($T_p=0.4$), a hint on possible large finite size effects in particle based simulations.

In EPM, the SB dynamics share some similarity and significant differences. We report the activity profile $P(x)$ in In Fig.~\ref{fig:sbgrowth}(c) for $L=128$ which corresponds to a system about $3$ times larger than our particle simulations. This choice will be motivated below. As in atomistic simulations, we find a progressive invasion of the band in the sample and a larger width at a given strain for the most poorly annealed sample. Importantly, we find that some lattice artifacts come into play for a small $L$. For example, if we prepare a very stable glass and use a lattice that is aligned along the imposed shear direction, here simple shear, we observe realizations where the SB is pinned and do not grow even for extremely large strains, see triangles in Fig.~\ref{fig:sbgrowth}(d) for $L=40$. Here, the positive interference of STZ within one slip line unloads the surrounding blocks outside the SB. This artifact disappears for larger disorder or larger grids, where the system always finds weak regions to propagate the band in the transverse direction. Moreover, we speculate that this artifact will disappear in a more refined vectorial EPM where blocks can be activated even-though their slip line is not perfectly aligned with the maximum stress directions, which was shown to be the case in particle based simulations~\cite{xu2021atomic}. In the thermodynamic limit $L\to \infty$, we find the SB growth to be a diffusive process with $w\sim \sqrt{\gamma}$, see Fig.~\ref{fig:sbgrowth}(d). Understanding the discrepancy between atomistic simulations and our mesoscale model for the aging of shearband is left for future investigations and could serve to formulate more refined lattice models.

\section{Discussion and conclusion}
\label{sec:discussion}

In this paper, we have developed a new method to probe mechanical heterogeneities in amorphous solids. We have derived expressions for the non-linear micromechanics of local pairwise potential energy and stresses that include both the affine and non-affine deformation. Firstly, we have measured the statistics for the coefficients of a non-linear strain expansion of the energy in glasses featuring a wide range of mechanical disorders in both 2D and 3D. We have highlighted the presence of anomalously large values of pairwise strain derivatives that we have linked to the presence of soft quasilocalized modes that promote large non-affinities. We have derived their expected asymptotic scaling based on the universal quartic law of the non-phononic density of states $D(\omega)\sim\omega^4$. Furthermore, we have shown that the distribution of pairwise derivatives allows one to extract the range of the depletion of soft modes as a function of our control parameter, here the parent temperature $T_p$ from which our glasses were first equilibrated before a final quench to zero temperature. As our formalism only requires solving linear equations, our method can be used to check glass stability in an efficient manner for both 2D and 3D glasses.

In a second part, we have shown that our formalism can be applied in the same spirit as the "frozen matrix" to extract stress activation threshold. We have highlighted the role of the preparation protocol on the depletion of regions having a small local yield stress. Moreover, we have quantitatively compared the yield stress distribution $P(\sigma_Y)$ obtained by our bond micromechanics and the original "frozen matrix". We have found that our method allows one to extract a good estimate of the average yield stress of a material solely from the as-cast configuration. However, the bond micromechanics overestimate large stress activation leading to a large variance of $P(\sigma_Y)$ compared with the "frozen matrix" method. In the future, we could improve local yield values by incorporating a semi-rigid matrix in our implementation, such as recently proposed in Refs.~\cite{adhikari2023soft,rottler2023thawed}.

Next, we have used local yield rules to parameterize a simple 2D scalar elasto-plastic model (EPM). We have shown that one can well predict the stress-strain response of materials exhibiting a wide range of glass stability that covers both a continuous homogeneous ductile flow and a discontinuous yielding transition with the formation of a shearband. Here, our only fit parameter was to set a fixed amplitude for the eigenstrain $\epsilon_{\rm pl}$ of a local shear transformation. We have shown that our mesoscopic model can recover the stress-strain response of material exhibiting a wide range of initial mechanical disorder, as well as size dependence stress fluctuations in the steady state. One possible improvement in the future will be to directly extract estimates of $\epsilon_{\rm pl}$ using the detection of non-linear modes~\cite{richard2023detecting} and fit procedures~\cite{albaret2016mapping,nicolas2018orientation} or the contour integral method developed in Ref.~\cite{moriel2020extracting}. This could allow for a complete calibration of a tensorial mesoscale model.

Finally, we highlighted some differences between our atomistic simulations and our EPM. In particular, we have monitored the aging dynamics of shearband and have found a subdiffusive and diffusive growth for particle based simulations and our elasto-plastic models, respectively. This confirms some earlier observations in 3D simulations~\cite{golkia2020flow}. Understanding missing ingredients in EPMs to capture such a slow growth will be valuable to refine future mesoscale models of plasticity in amorphous solids and to study fatigue failure such as during a cyclic loading~\cite{parmar2019strain,Khirallah2021yielding,kumar2022mapping,liu2022fate}.

{\bf Acknowledgments} We warmly thank K. Martens, S. Patinet, and E. Lerner for fruitful discussions and comments. D.R.~acknowledges support by the H2020-MSCA-IF-2020 project ToughMG  (No.~101024057) and computational resources provided by GENCI (No.~AD010913428).\\

\appendix

\section{Model and protocol}

The computer glass model we employed is a slightly modified variant of the model put forward in \cite{ninarello2017models}. We enclose $N$ particles of equal mass $m$ in a square box of volume $V\!=\! L^3$ with periodic boundary conditions, and associate a size parameter $\lambda_i$ to each particle, drawn from a distribution $p(\lambda)\!\sim\!\lambda^{-3}$. We only allow $\lambda_{\mbox{\tiny min}}\!\le\!\lambda_i\le\!\lambda_{\mbox{\tiny max}}$ with $\lambdabar\!\equiv\!\lambda_{\mbox{\tiny min}}$ forming our units of length, and $\lambda_{\mbox{\tiny max}}\!=\!2.22\lambdabar$. The number density $N/V\!=\!0.86\lambdabar^{-3}$ (in 2D) and $N/V\!=\!0.58\lambdabar^{-3}$ (in 3D) are kept fixed. Pairs of particles interact via the the inverse-power law pairwise potential
\begin{equation}
\varphi_{\mbox{\tiny IPL}}(r_{ij}) = \varepsilon\left( \sFrac{\lambda_{ij}}{r_{ij}} \right)^\beta\,,
\end{equation}
where $\varepsilon$ is a microscopic energy scale. Distances in this model are measured in terms of the interaction lengthscale $\lambda$ between two `small' particles, and the rest are chosen to be $\lambda_{ij}\!=\!1.18\lambda$ for one `small' and one `large' particle, and $\lambda_{ij}\!=\!1.4\lambda$ for two `large' particles. In finite systems under periodic boundary conditions, a variant of the IPL model with a finite interaction range should be employed, otherwise the potential is discontinuous due to the periodic boundary conditions. We chose the form
\begin{equation}
\varphi_{\mbox{\tiny IPL}}(r_{ij}) = \left\{ \begin{array}{ccc}\varepsilon\left[ \left( \sFrac{\lambda_{ij}}{r_{ij}} \right)^\beta + \sum\limits_{\ell=0}^q c_{2\ell}\left(\sFrac{r_{ij}}{\lambda_{ij}}\right)^{2\ell}\right]&,&\sFrac{r_{ij}}{\lambda_{ij}}\le x_c\\0&,&\sFrac{r_{ij}}{\lambda_{ij}}> x_c\end{array} \right.\!\!,
\end{equation}
where $x_c$ is the dimensionless distance for which $\varphi_{\mbox{\tiny IPL}}$ vanishes continuously up to $q$ derivatives. The coefficients $c_{2\ell}$, determined by demanding that $\varphi$ vanishes continuously up to $q$ derivatives, are given by
\begin{equation}
c_{2\ell} = \frac{(-1)^{\ell+1}}{(2q\!-\!2\ell)!!(2\ell)!!}\frac{(\beta\!+\!2q)!!}{(\beta\!-\!2)!!(\beta\!+\!2\ell)}x_c^{-(\beta+2\ell)}\,.
\end{equation}
We chose the parameters $x_c\!=\!1.4, n\!=\!10$, and $q\!=\!3$. The pairwise length parameters $\lambda_{ij}$ are given by 
\begin{equation}
\lambda_{ij} = \sFrac{1}{2}(\lambda_i + \lambda_j)(1 - n_a|\lambda_i - \lambda_j|)\,.
\end{equation}
Following \cite{ninarello2017models} we set the non-additivity parameter $n_a\!=\!0.1$. In what follows energy is expressed in terms of $\varepsilon$, temperature is expressed in terms of $\varepsilon/k_B$ with $k_B$ the Boltzmann constant, stress, pressure, and elastic moduli are expressed in terms of $\varepsilon/\lambdabar^3$.

Solids were created by first equilibrating a melt at a parent temperature $T_p$ using SWAP Monte Carlo~\cite{ninarello2017models}, followed by an energy minimization using a standard nonlinear conjugate gradient algorithm. Both 2D and 3D systems have an onset temperature around $T\simeq0.6$.

\section{Notations and formalism}

Here, we provide useful notations for evaluating high order derivatives of the potential energy $U$ for pairwise interactions (e.g. $\mathcal{H}$, $\mathcal{T}$, ...) as well as expressions for contraction of those tensors with displacement field ($\mathcal{H}\doubleCdot\vv\uv$). For convenience, we first introduce scalar notations, which can be used to simplified tensor contractions:
\begin{equation}
\text{(first)}=\frac{\varphi_\iota^{'}}{r_\iota},
\end{equation}
\begin{equation}
\text{(second)}=\frac{\varphi_\iota^{''}}{r_\iota^2}-\frac{\varphi_\iota^{'}}{r_\iota^3},
\end{equation}
\begin{equation}
\text{(third)}=\frac{\varphi_\iota^{'''}}{r_\iota^3}-3\frac{\varphi_\iota^{''}}{r_\iota^4}+3\frac{\varphi_\iota^{'}}{r_\iota^5},
\end{equation}
\begin{equation}
\text{(fourth)}=\frac{\varphi_\iota^{''''}}{r_\iota^4}-6\frac{\varphi_\iota^{'''}}{r_\iota^5}+15\frac{\varphi_\iota^{''}}{r_\iota^6}-15\frac{\varphi_\iota^{'}}{r_\iota^7},
\end{equation}
\begin{equation}
\text{(fifth)}=\frac{\varphi_\iota^{'''''}}{r_\iota^5}-10\frac{\varphi_\iota^{''''}}{r_\iota^6}+45\frac{\varphi_\iota^{'''}}{r_\iota^7}-105\frac{\varphi_\iota^{''}}{r_\iota^8}+105\frac{\varphi_\iota^{'}}{r_\iota^9},
\end{equation}

\subsection{Energy tensor contractions}
The complete set of contractions used throughout this paper read:
\begin{itemize}
\item Jacobian:
\begin{equation}
\calBold{J}\cdot\vv = \sum\text{(first)}(\xv\vv)
\end{equation}
\item Hessian:
\begin{eqnarray}
\calBold{H}\cdot\vv &=& \sum\text{(second)}(\xv\vv)\xv \nonumber\\
&+&\sum\text{(first)}\vv
\end{eqnarray}
\begin{eqnarray}
\calBold{H}\doubleCdot\vv\uv &=& \sum\text{(second)}(\xv\vv)(\xv\uv) \nonumber\\
&+&\sum\text{(first)}(\vv\uv)
\end{eqnarray}
\item Tessian:
\begin{eqnarray}
\calBold{T}\tripleCdot\vv\uv &=& \sum\text{(third)}(\xv\vv)(\xv\uv)\xv\\
&+&\sum\text{(second)}[(\xv\vv)\uv+(\xv\uv)\vv+(\uv\vv)\xv]\nonumber
\end{eqnarray}
\begin{eqnarray}
\calBold{T}\tripleCdot\vv\uv\wv &=& \sum\text{(third)}(\xv\vv)(\xv\uv)(\xv\wv)\nonumber\\
&+&\sum\text{(second)}[(\xv\vv)(\uv\wv)+(\xv\uv)(\vv\wv)\nonumber\\
&+&(\uv\vv)(\xv\wv)]
\end{eqnarray}
\item Messian:
\begin{eqnarray}
\calBold{M}\quadCdot\vv\uv\wv &=& \sum\text{(fourth)}(\xv\vv)(\xv\uv)(\xv\wv)(\xv)\nonumber\\
&+&\sum\text{(third)}[(\xv\uv)(\xv\wv)(\vv)+(\xv\vv)(\xv\wv)(\uv)\nonumber\\
&+&(\xv\vv)(\xv\uv)(\wv)+(\xv\vv)(\uv\wv)(\xv)\nonumber\\
&+&(\xv\uv)(\vv\wv)(\xv)+(\xv\wv)(\vv\uv)(\xv)]\nonumber\\
&+&\sum\text{(second)}[(\uv\vv)(\wv)+(\uv\wv)(\vv)\nonumber\\
&+&(\vv\wv)(\uv)]
\end{eqnarray}
\begin{eqnarray}
\calBold{M}\quadCdot\vv\uv\wv\zv &=& \sum\text{(fourth)}(\xv\vv)(\xv\uv)(\xv\wv)(\xv\zv)\nonumber\\
&+&\sum\text{(third)}[(\xv\uv)(\xv\wv)(\vv\zv)+(\xv\vv)(\xv\wv)(\uv\zv)\nonumber\\
&+&(\xv\vv)(\xv\uv)(\wv\zv)+(\xv\vv)(\uv\wv)(\xv\zv)\nonumber\\
&+&(\xv\uv)(\vv\wv)(\xv\zv)+(\xv\wv)(\vv\uv)(\xv\zv)]\nonumber\\
&+&\sum\text{(second)}[(\uv\vv)(\wv\zv)+(\uv\wv)(\vv\zv)\nonumber\\
&+&(\vv\wv)(\uv\zv)]
\end{eqnarray}
\end{itemize}
Note that for simplicity pairwise indexes and scalar products are omitted, i.e, $\sum \xv=\sum_\iota\xv_\iota$ and $\vv\uv$ stands for the scalar product $\vv\cdot\uv$.

\subsection{Stress expansion contractions}

Below we provide the necessary contractions needed for the non-linear strain expansion of the stress observable ${\cal O} = \calBold{J}\cdot (\calBold{B}\cdot \xv)$. The first, second, and third derivatives with respect to particle coordinates $\xv$ and contractions with displacement fields $\uv$, $\vv$, and $\wv$ reads:
\begin{eqnarray}
\frac{\partial {\cal O}}{\partial\xv}\doubleCdot \uv = \calBold{H}\doubleCdot (\calBold{B}\cdot \xv)\uv + \calBold{J}\cdot (\calBold{B}\cdot \uv),
\end{eqnarray}
\begin{eqnarray}
\frac{\partial^2 {\cal O}}{\partial\xv\partial\xv}\doubleCdot \uv\vv &=& \calBold{T}\tripleCdot (\calBold{B}\cdot \xv)\uv\vv+\calBold{H}\doubleCdot (\calBold{B}\cdot \vv)\uv\nonumber\\
&+&\calBold{H}\doubleCdot (\calBold{B}\cdot \uv)\vv,
\end{eqnarray}
and 
\begin{eqnarray}
\frac{\partial^3 {\cal O}}{\partial\xv\partial\xv\partial\xv}\tripleCdot \uv\vv\wv &=& \calBold{M}\quadCdot (\calBold{B}\cdot \xv)\uv\vv\wv + \calBold{T}\tripleCdot (\calBold{B}\cdot \wv)\uv\vv \nonumber\\
&+& \calBold{T}\tripleCdot (\calBold{B}\cdot \vv)\uv\wv\nonumber\\
&+& \calBold{T}\tripleCdot (\calBold{B}\cdot \uv)\vv\wv.
\end{eqnarray}

\section{Linear and non-linear strain dynamics for a solid under stress}

In this section, we provide a detailed explanation on how our formalism can be used to compute the linear and non-linear stress-strain dynamics for a solid that exhibit a non-zero stress tensor. Without loss of generality, we will consider the dynamics of the shear stress $\sigma_{xy}$ as well as the pressure $P=-\Tr(\calBold{\sigma})/\dbar$. Following our formalism, these two quantities can be computed by defining the tensor $\calBold{B}_{xy}=\hat{x}\hat{y}$ and $\calBold{B}_{P}=\calBold{I}$, where $\calBold{I}$ is the identity matrix, giving $\sigma_{xy}=\calBold{J}\cdot (\calBold{B}_{xy}\cdot \xv)/V$ and $P=-\calBold{J}\cdot (\calBold{B}_{P}\cdot \xv)/(V\dbar)$. We additionally define the operator $\mathcal{O}_{xy}=\calBold{J}\cdot (\calBold{B}_{xy}\cdot \xv)$ and $\mathcal{O}_P=-\calBold{J}\cdot (\calBold{B}_{P}\cdot \xv)$.

%%%%%%%%%%%%%%%%%%%%%%%%%%%%%%%%%%%%%%%%%%%%%%%%%%%%%%%%%%%%%%%%%%%%%%%
\begin{figure}[t!]
  \includegraphics[width = 0.5\textwidth]{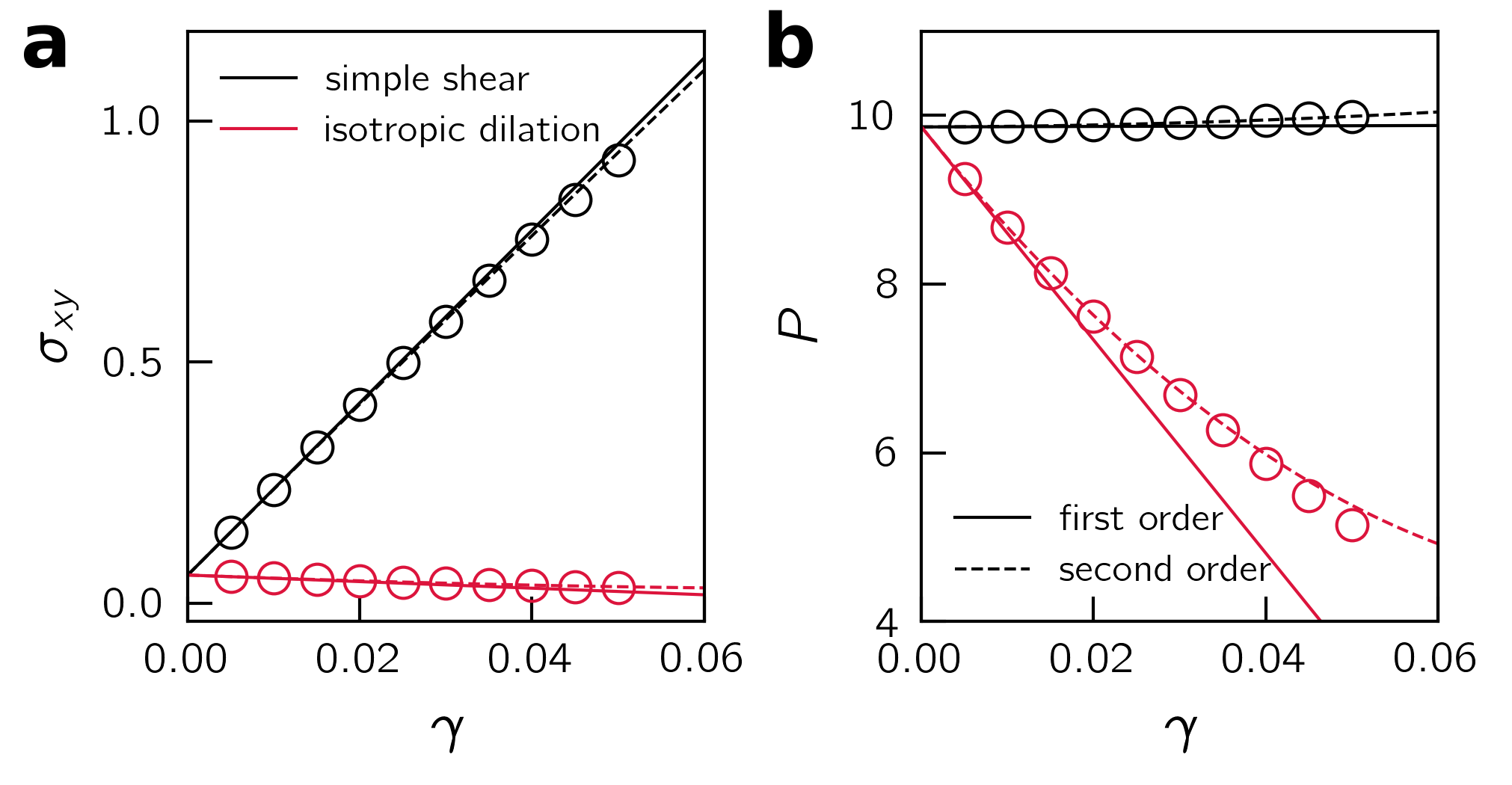}
  \caption{\footnotesize (a) Shear stress $\sigma_{xy}$ versus the strain $\gamma$ for a simple shear (black) and isotropic dilation (red). The empty circles are the actual AQS dynamics. The solid and dashed lines are the linear and nonlinear (up to $\gamma^2$) analytical prediction, respectively. (b) Same as in (a) but for the pressure $P=-\Tr(\calBold{\sigma})/\dbar$.}
  \label{fig:geometry}
\end{figure}
%%%%%%%%%%%%%%%%%%%%%%%%%%%%%%%%%%%%%%%%%%%%%%%%%%%%%%%%%%%%%%%%%%%%%%%

Now let us consider an arbitrary deformation $\mathcal{D}(\gamma)$ that does not necessarily conserve the volume of the sample, i.e., $dV(\gamma)/d\gamma\ne 0$. When taking total derivative $d/d\gamma$ on either $\sigma_{xy}$ or $P$, we need to evaluate derivatives of $d^n V^{-1}(\gamma)/d\gamma^n$ on top of derivatives of $\mathcal{O}_{xy}$ and $\mathcal{O}_{P}$. As an example, for an isotropic dilation (cf. Eq.~\ref{eq:dilation}) with strain $\gamma$, one finds $dV^{-1}/d\gamma|_0=-\dbar/V_0$ and $d^2V^{-1}/d\gamma^2|_0=\dbar^2/V_0$, with the reference volume $V_0$. Thus, the linear and first non-linear coefficients of the pressure dynamics can be expressed as
\begin{equation}
\frac{d P}{d\gamma}\Bigg|_0=-\dbar P + \frac{1}{\dbar V_0}\frac{d \mathcal{O}_{P}}{d\gamma},
\end{equation}
and
\begin{equation}
\frac{d^2 P}{d\gamma^2}\Bigg|_0=\dbar^2 P - \frac{2}{V_0}\frac{d \mathcal{O}_{P}}{d\gamma}+\frac{1}{\dbar V_0}\frac{d^2 \mathcal{O}_{P}}{d\gamma^2},
\end{equation}
respectively. To evaluate $d \mathcal{O}_{P}/d\gamma$ and $d^2 \mathcal{O}_{P}/d\gamma^2$, one simply need to use the formalism described in the main text with $\mathcal{A}=\calBold{I}$. Note that the first term of these expressions vanishes for an unstressed solid. The shear stress dynamics follows easily from the same methodology. Below, we compare our analytical predictions with the true AQS strain dynamics of the shear stress (Fig.~\ref{fig:geometry}(a)) and pressure (Fig.~\ref{fig:geometry}(b)) for both a simple shear deformation and an isotropic dilation of a stable glass up to $\gamma=5\%$. We find a perfect agreement between our theory and AQS. Deviations are to be expected for large deformations as we can see from comparing the linear response (solid line) and the non-linear (dashed lines) dynamics of order $\mathcal{O}(\gamma^2)$.

\section{Shearband nucleation: finite size study}

\begin{figure}[h!]
  \includegraphics[width = 0.5\textwidth]{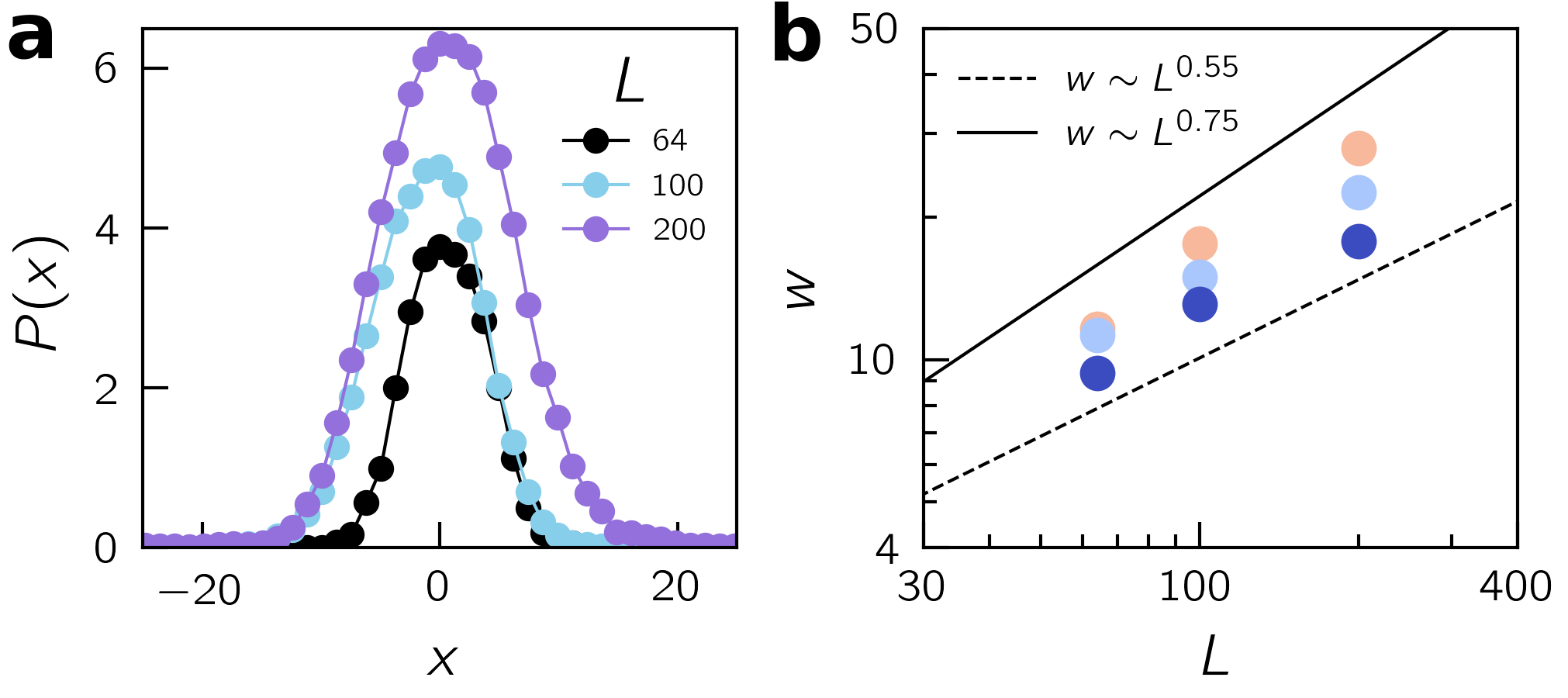}
  \caption{\footnotesize (a) Plastic profile measured after the yielding transition at $\gamma=0.15$ for different system sizes and $T_p=0.2$. (b) Shearband width $w$ versus the box length $L$ for different parent temperatures. The color code is the same as in Fig.~\ref{fig:sbgrowth}.}
  \label{fig:sbfinite}
\end{figure}

Here, we provide additional data that quantify finite size effect in the plastic profile associated with the nucleating shearband post yielding. In Fig.~\ref{fig:sbfinite}(a), we show the palstic profile $P(x)$ measured at $\gamma=0.15$ for different system sizes. We find that the width of the shearband grows with the box length $L$. In Fig.~\ref{fig:sbfinite}(b), we report $w$ as a function of the box length $L$ for the same 3 different parent temperatures as reported in Fig.~\ref{fig:sbgrowth}. We find a subextensive growth according to $w\sim L^\alpha$, with $\alpha$ ranging from $0.55$ to $0.75$. From the limited range of box size $L$ available in particle based simulations, it is rather unclear if $\alpha$ does depend on the initial degree of mechanical disorder.

%%%%%%%%%%%%%%%%%%%%%%%%%%%%%%%%%%%%%%%%%%%%%%%%%%%%%%%%%%%%%%%%%%%%%%%
%%%%%%%%%%%%%%%%%%%%%%%%%%%%%%%%%%%%%%%%%%%%%%%%%%%%%%%%%%%%%%%%%%%%%%%

%\bibliography{reference}

\end{document}